\documentclass[aps,prd,twocolumn,groupedaddress, showkeys, showpacs]{revtex4}
\pdfoutput=1
\usepackage{graphicx,subfigure,enumerate,pstricks,latexsym}
\usepackage[english]{babel}

\usepackage{color}
\usepackage{babel}
\usepackage{amsmath}
\usepackage{amssymb}
\usepackage{graphicx,subfigure}
\usepackage{esint}
\usepackage{longtable}
\usepackage{dcolumn}


\newcommand{\dd}{{\rm d}}
\newcommand{\Dr}{\Delta_r}

\newcommand{\LL}{{\cal L}}
\newcommand{\JJ}{{\cal J}}
\newcommand{\EE}{{\cal E}}
\newcommand{\BB}{{\cal B}}

\def\beq{\begin{equation}}
\def\eeq{\end{equation}}
\def\bea{\begin{eqnarray}}
\def\eea{\end{eqnarray}}
\providecommand{\dif}{\mathrm{d}} \def\d{\dif} 

\begin{document}

\title{Acceleration of particles in spacetimes of black string}

\author{Arman Tursunov$^{1,2,3}$}
 \email{arman@astrin.uz}

\author{Martin Kolo\v{s}$^{3}$}
 \email{martin.kolos@fpf.slu.cz}

\author{Ahmadjon~Abdujabbarov$^{1,2}$}
 \email{ahmadjon@astrin.uz}

\author{Bobomurat Ahmedov$^{1,2}$}
 \email{ahmedov@astrin.uz}

 \author{Zden\v{e}k Stuchl{\'i}k$^{3}$}
 \email{zdenek.stuchlik@fpf.slu.cz}

 \affiliation{%
$^1$ Institute of Nuclear Physics, Ulughbek, Tashkent 100214, Uzbekistan\\
$^2$ Ulugh Begh Astronomical Institute, Astronomicheskaya 33, Tashkent 100052, Uzbekistan\\
$^3$ Institute of Physics, Faculty of Philosophy and Science, Silesian University in Opava, \\
Bezru{\v c}ovo n{\'a}m.13, CZ-74601 Opava, Czech Republic }

\begin{abstract}

{The particle acceleration mechanism in $S^2 \times\mathbb{R}^1$
topology, namely, in the spacetime of the five-dimensional compact
black string, has been studied.
The expression of
center-of-mass energy of the colliding neutral particles near
static black string has been found.
The collision of a charged particle moving at the innermost stable
circular orbit with a neutral particle coming from infinity has
been considered when black string is immersed in external uniform
magnetic field.
It {has been} shown that the unlimited center-of-mass energy can
be approached in the case of the extremal rotation of the black
string which is similar to the analogous effect in Kerr spacetime.
We have also {obtained} that the scattering energy of particles in
the center-of-mass system can take arbitrarily large values not
only for extremal black string but also for the nonextremal one.
{It has been derived} that the presence of the extra dimension
can, in principle, increase the upper limit of efficiency of
energy extraction from the extremely rotating black string  up  to
203\% versus 143\% which can be extracted from the extreme Kerr
black hole.}

\end{abstract}


\maketitle

\section{Introduction}

String theories give an infinite series of corrections to the
theory of gravity and {consequently} more types of symmetries than
those commonly assumed in standard Einstein general relativity.
{The bigger} number of symmetries can be separated by the range of
horizon topologies. The event horizon of the four-dimensional
black hole is topologically spherical $S^2$, since the
cross-section of the event horizon is a two-sphere while for 5D
black holes the topology of the horizon is $S^3$. {However, for} a
spacetime with one extra dimension, which is compactified to the
circle with length $2\pi L$, the spacetime {has} $S^2
\times\mathbb{R}^1$ topology \cite{Bondarescu:2008}. This
one-dimensional extended object surrounded by a horizon we will
call \textit{black string} \cite{Horowitz2002}.

It has been recently
shown that new interesting properties of the geodesic motion occur
if one adds the extra dimension to Schwarzschild or Kerr
spacetimes \cite{Grunau:2013}. When the additional dimension $w$
is included to the well known four-dimensional Schwarzschild
spacetime metric, the obtained solution will give us the spacetime
metric of a five-dimensional static black string. By the same way
one may obtain the spacetime metric for the rotating black string
adding extra dimension to the spacetime of Kerr black hole
\cite{Horowitz2002}.

One can use the following notations:
\begin{equation}
X^{\alpha} = (x^a,w), \quad \alpha=0,1,2,3,4, \quad a=0,1,2,3
\end{equation}
for the complete set of coordinates which cover the spacetime. The
coordinates $x^\mu$ form the {four-dimensional} spacetime with
metric $g_{\mu\nu} = (-,+,+,+)$. Then, the spacetime metric of the
black string living in five dimensions takes the following form:
\begin{equation}
ds^2 = g_{\mu\nu} dx^\mu dx^\nu + dw^2,
\end{equation}
where $g_{\mu\nu} dx^\mu dx^\nu$ can be a metric of any
{four-dimensional} black hole spacetime.

The existence of higher dimensions, which are mostly assumed to be
compact, is essential for
the intrinsic congruency of the field theory.
Charged rotating black holes are
considered in~\cite{aliev1}, while optical phenomena in the field of rotating black hole in braneworld has been studied in~\cite{schee1}.
Equatorial circular orbits and the
motion of the shell of dust in the field of a rotating naked singularity have been studied in detail in~\cite{zd80}
%


{ Observational possibilities of testing the braneworld black hole
models at an astrophysical scale have been intensively discussed
in the literature during the last several years, for example,
through the gravitational lensing~\cite{pk08}, the motion of test
particles~\cite{aaprd,zdnk,zdnkktrlv}, and the classical tests of
general relativity (perihelion precession, deflection of light,
and the radar echo delay) in the Solar System (see, {e.g.},
Ref.~\cite{lobo08}). The energy flux, the emission spectrum, and
accretion efficiency from the accretion disks around several
classes of static and rotating braneworld black holes have been
obtained in~\cite{pkh08}. The complete set of analytical solutions
of the geodesic equation of massive test particles in higher
dimensional spacetimes which can be applied to braneworld models
is provided in the recent paper \cite{Lam08}.}

The geodesic motion of test particles in the spacetimes of static
and rotating black strings and {in the various spacetimes related
to cosmic strings} have been studied in detail in the recent paper
\cite{Grunau:2013} and in
\cite{Aliev:1988wv,Galtsov:1989ct,Chakraborty:1991mb,Ozdemir:2003km,Ozdemir:2004ne}.
The dynamics of a test particle in the spacetimes of Schwarzschild
and Kerr black holes pierced by string has been studied in papers
\cite{Hackmann:2009rp,Hackmann:2010ir}, which was mentioned in
\cite{Grunau:2013}. Moreover, the solutions of the dynamical
equations in the gravitational field of cosmic strings, such as
Abelian-Higgs strings \cite{Hartmann:2010rr}, two interacting
Abelian-Higgs strings \cite{Hartmann:2012pj}, and cosmic
superstrings \cite{Hartmann:2010vp} were mentioned in the paper
\cite{Grunau:2013}.\\

It has been shown by Ba$\mathrm{\tilde{n}}$ados, Silk and West
(BSW) \cite{BSW:2009} that a rotating {axially symmetric} black
hole {can} act as a particle accelerator to arbitrarily high
energies in the center-of-mass frame of the collision of a pair of
particles. In particular, the BSW effect takes place when
particles have the properly chosen values of angular momentum.
Nowadays, the effect of infinite energy in the center-of-mass
frame due to the collision of particles attracts much attention;
see, e.g., \cite{Said-Adami:2011}-\cite{Abd-Ahm-Jur:2013}. It
seems that this effect has a quite general character. The
acceleration of particles by a spinning black hole
\cite{Jac-Sot:2010}, cylindrical black hole
\cite{Said-Adami:2011}, weakly magnetized black hole
\cite{Frolov:2012}, black hole with gravitomagnetic charge
\cite{Abd-Tur-Ahm-Kuv:2013}, or Kerr naked singularity
\cite{Pat-Josh:CQG:2011} has been analyzed regarding the
possibility of the production of the particles with unlimited
energies. Energetic processes in the superspinning Kerr spacetimes
have been also
studied in papers
\cite{Stu-Hle-Tru:CQG:2011,Stuchlik-Schee:CQG:2012-1,Stuchlik-Schee:CQG:2012-2}.
By constructing escape null cones, it has been explicitly
demonstrated that the high energy collisions occurring in the
field of near extreme Kerr superspinars can be directly observed
by distant observers~\cite{Stuchlik-Schee:CQG:2013}.
Optical phenomena related to the appearance of Keplerian accretion
discs orbiting Kerr superspinars have been studied in
Ref.~\cite{stuchlik_10}
Black hole production from ultrarelativistic collisions is
explored in Ref.~\cite{Ata-Ahm-Sha:2013,takami}.
Collision of particles in different  trajectories are considered in~\cite{japan1,japan2}.
High energy collision  of particles in the vicinity of black holes in  higher dimensions have been studied in Ref.~\cite{japan3}.
Efficiency of the particle collision and upper limit of energy extraction for Kerr black hole through particle acceleration are studied in~\cite{japan4,japan5}.
The particle acceleration around 5 dimensional rotating black hole in supergravity theory has been considered in recent paper~\cite{dadhich}

\begin{figure}[h]
 \centering
   \includegraphics[width=6cm]{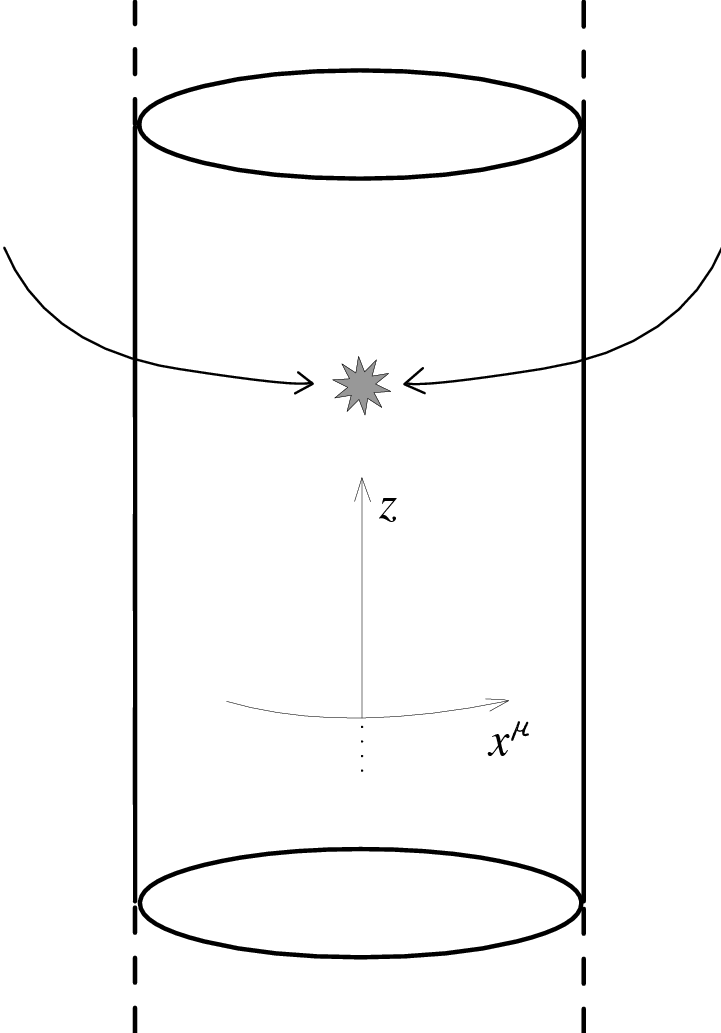}
 \caption{Schematic picture of two particles colliding near the event horizon of a black string.} \label{schema}
\end{figure}

In the present paper, {our} aim is to show that a similar effect
of particle collision with high center-of-mass energy is also
possible when the string theory phenomena {are included}. Namely,
in the simplest case {when} a space dimension is added to the
four-dimensional black hole solutions, it can be represented as
the spacetime of black string (see, Fig.\ref{schema})

The paper is organized as follows: in Sec.~\ref{secII} the test
particles motion as well as their acceleration near non-rotating
5D black string is studied. The Section~\ref{secIII} is devoted to
{investigation of} charged particle motion and acceleration of
particles around black string in external magnetic field. The
rotating black string as particle accelerator is {explored} in
Sec.~\ref{secIV}. The energy extraction efficiency is studied in
{subsequent} Sec.~\ref{secV}. We conclude our results in
Sec.~\ref{secVI}.

Throughout the paper, we use a plus signature and a system of
geometric units in which $G = 1 = c$. However, for those
expressions with an astrophysical application, we have written the
speed of light explicitly.

\section{Black string as accelerator of neutral particles}
\label{secII}

\subsection{Dynamical equations}

Now, the geodesic motion of a test neutral particle in the
spacetime of nonrotating black string will be studied. In
addition, we will find the analytical solutions of the equations
of motion and show the representative trajectories of a test
particle in the vicinity of nonrotating black string.

When the additional dimension $w$ is included to the metric of the
Schwarzschild spacetime, the obtained solution takes the following
form~\cite{Grunau:2013}
\begin{eqnarray}
 {\rm d}s^2&=&-\left(1-\frac{2M}{r}\right){\rm d}t^2 + \left(1-\frac{2M}{r}\right)^{-1}{\rm d}r^2 \nonumber\\
 &&+ r^2({\rm d}\vartheta^2
  + \sin^2\vartheta{\rm d}\varphi^2) +{\rm d}w^2 \, ,
  \label{statmetric}
\end{eqnarray}
{where $M$ is the mass density of the black string or the mass per
unit length}. The solution \ref{statmetric} is the solution of a
neutral uniform black string spacetime and the singularity is
placed at the position $r = 0$. The event horizon is at the
position $r_{\rm H} = 2 M$.

{Consider the geodesic motion of a neutral point particle in the
spacetime of the nonrotating static black string given by the
expression (\ref{statmetric}). In order to find the trajectory of
a particle one can use the Hamilton-Jacobi equation in the general
form \cite{Grunau:2013}}
%
\begin{equation}
 \frac{\partial S}{\partial \lambda} + \frac12 g^{\mu\nu} \frac{\partial S}{\partial x^\mu}\frac{\partial S}{\partial
 x^\nu}=0,
\label{eqn:ham-jac}
\end{equation}
where we parameterize by $\lambda$ being an affine parameter of
the geodesic line of the test particle.

{The Hamiltonian of the test particle then can be written in the
form
\begin{equation}
2 H = g^{a b}_{(4)} P_{ a} P_{ b} + P_{w}^2 +
 m^2,
\end{equation}
where the momenta of the particle are associated with the first
derivative of the Hamilton-Jacobi action, $S$, with respect to the
corresponding coordinate as $P_{\mu} = \partial S / \partial
x^{\mu}$. The term $g^{a b}_{(4)}$ denotes the four-dimensional
part of the space-time metric (\ref{statmetric}). }

{Due to the symmetries of the background spacetime of a
nonrotating black string, there are three constants of motion for
each particle, whuch means there are three conserved quantities
for any geodesic motion, namely, the energy $E$, the angular
momentum $L$, and a new constant of motion $J$, which is related
to the extra dimension $w$. The constants of the motion are
related to the five-momentum as follows}:
\begin{eqnarray}
P_t=-E, \ \ P_{\varphi}=L, \ \ P_w=J.
\end{eqnarray}
{One may look for the separable solutions of the Hamilton-Jacobi
equations decomposing the action in the form \cite{Grunau:2013} }
\begin{equation}
 S=\frac12 m_0^{2} \lambda -Et+L\varphi + Jw + S_r(r) \, .
 \label{actionSBS}
\end{equation}

{To study the motion of test particles in the black string
spacetime it is convenient to use the fact that the entire
trajectory must lie on the plane if the initial position and the
tangent vector to the trajectory of the particle lie on a plane
that contains the center of the gravitating body. Without loss of
generality we may therefore restrict ourselves to the study of
equatorial trajectories with $\vartheta={\pi}/{2}$. Hereafter, we
set $\vartheta={\pi}/{2}$, since the orbits always lie in a plane
which can be selected as equatorial one.} {Substituting
(\ref{actionSBS}) into (\ref{eqn:ham-jac}) and using
(\ref{statmetric}), one can find the radial part of the action
(\ref{actionSBS}) in the form
\bea
S_r &=& \int \Big[\left(1-\frac{2M}{r}\right)^{-2} E^2  \nonumber \\
&& -\left(1-\frac{2M}{r}\right)^{-1}
\left(\frac{L^2}{r^2}+J^2+m^2\right)\Big]^{1/2} \d r
\eea
}

Let us denote for convenience \cite{Grunau:2013}
\begin{eqnarray}
{\cal E} = \frac{E}{m}, \qquad {\cal L} = \frac{L}{m}, \qquad
{\cal J} = \frac{J}{m},
\end{eqnarray}
where ${\cal E}$, ${\cal L}$, and ${\cal J}$ are the specific
energy, angular momentum and new constant of motion per unit of
mass $m$ of the particle.

Then the Hamilton-Jacobi equation \eqref{eqn:ham-jac} can be
separated and gives the equations of motion in differential form
for each component as
\begin{eqnarray}
 \frac{d t}{d\tau} &=& {\cal E}\left(1-\frac{2M}{r}\right)^{-1}, \label{eqn:t-equation}\\
 \frac{d \varphi}{d \tau} &=& \frac{{\cal L}}{r^2}, \label{eqn:phi-equation}\\
 \frac{d w}{d \tau} &=& {\cal J}, \label{eqn:w-equation}\\
 \left(\frac{d r}{d \tau}\right)^2 &=& {\cal E}^2 - \left(1-\frac{2M}{r}\right)\left(1+\frac{{\cal
L}^2}{r^2}+{\cal J}^2\right). \label{eqn:r-equation}
\end{eqnarray}

The integral of motion $\cal{J}$ which is related to the symmetry
of the extra dimension $w$, appears in the $r$ and $w$ equations,
\eqref{eqn:r-equation} and \eqref{eqn:w-equation}, respectively,
while remaining equations do not change their original form
presented in the four-dimensional Schwarzschild spacetime
\cite{Grunau:2013}. The equations of motion
(\ref{eqn:t-equation})-(\ref{eqn:r-equation}) are invariant under
the following reversals of signs
\begin{equation}
\cal{L} \rightarrow -\cal{L}, \qquad \varphi\rightarrow -\varphi,
\end{equation}
and
\begin{equation}
{\cal J}\rightarrow -{\cal J},  \qquad  w \rightarrow -w,
\end{equation}

\subsection{Effective potential of the motion}

The number of turning points and the form of the effective
potential allow us to define the possible orbits in the given
spacetime. In the nonrotating black string vicinity there exist:
terminating orbits (TO), for which the trajectory of a particle
ends in singularity, escape orbits (EO), where the trajectories
are open to infinity, and bound orbits (BO), where the
trajectories are closed \cite{Grunau:2013}. Depending on the type
of orbit, the possible trajectories cover all region between $[0,
\infty)$.


 \begin{figure*}[t!]
\begin{center}
   \includegraphics[width=5.5cm]{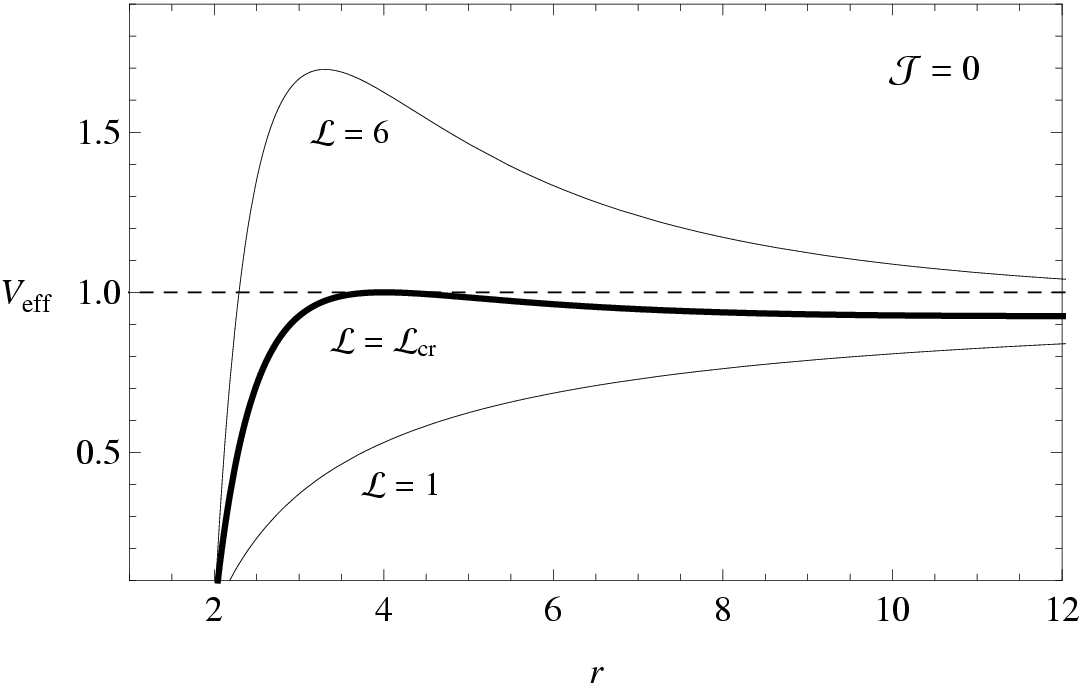} \label{potstata}
    \includegraphics[width=5.5cm]{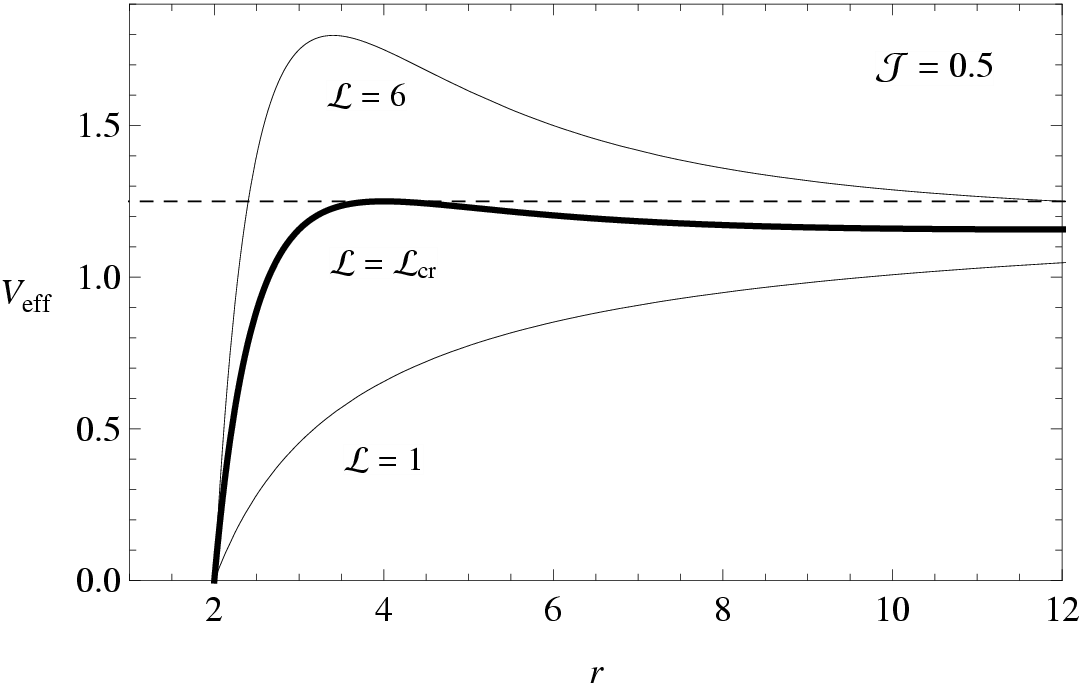} \label{potstatb}
     \includegraphics[width=5.5cm]{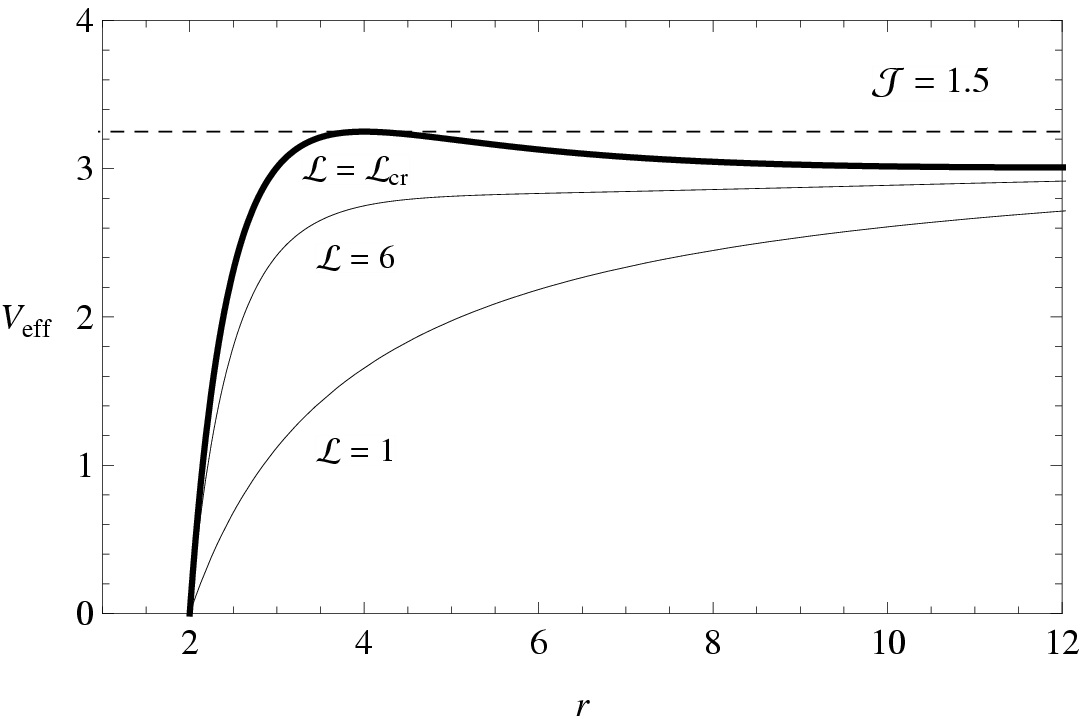} \label{potstatc}
  \caption{Effective potential of the motion for the different values of parameters $\LL$ and
  $\JJ$. Thick solid curves correspond to the particles with the
  critical angular momentum $\LL_{\rm cr}$ while dashed lines are
  the levels of the energy of a free particle at infinity. Thin solid curves are the potentials of the particles with the representative
  values of the angular momentum $\LL=1$ and $\LL=6$.}
 \label{pic:potential}
\end{center}
\end{figure*}

Let us introduce the effective potential $V_{\rm eff}$
\begin{equation}
 V_{\rm eff}=\left(1-\frac{2M}{r}\right)\left(1+\frac{{\cal L}^2}{r^2}+{\cal J}^2\right)
\end{equation}
which can be easily defined from Eq.~\eqref{eqn:r-equation} as
\begin{equation}
\left(\frac{dr}{d\tau}\right)^2={\cal E}^2 -V_{\rm eff}\, .
\label{eqn:turningpoints}
\end{equation}

In the asymptotical infinity $r\rightarrow\infty$, or in the flat
spacetime limit, the effective potential $V_{\rm eff}$ takes the
value $1+{\cal J}^2$. {In the opposite limiting case}
$r\rightarrow 0$, the effective potensial diverges, $V_{\rm eff}
\rightarrow -\infty$ \cite{Grunau:2013}. Equation
\eqref{eqn:turningpoints} also determines the turning points of an
orbit. The shapes of the effective potential for representative
values of $\LL$ and $\JJ$ are shown in Fig.\ref{pic:potential}.

\begin{figure*}
\subfigure[ \quad TO
$\LL=3$]{\includegraphics[width=0.3\hsize]{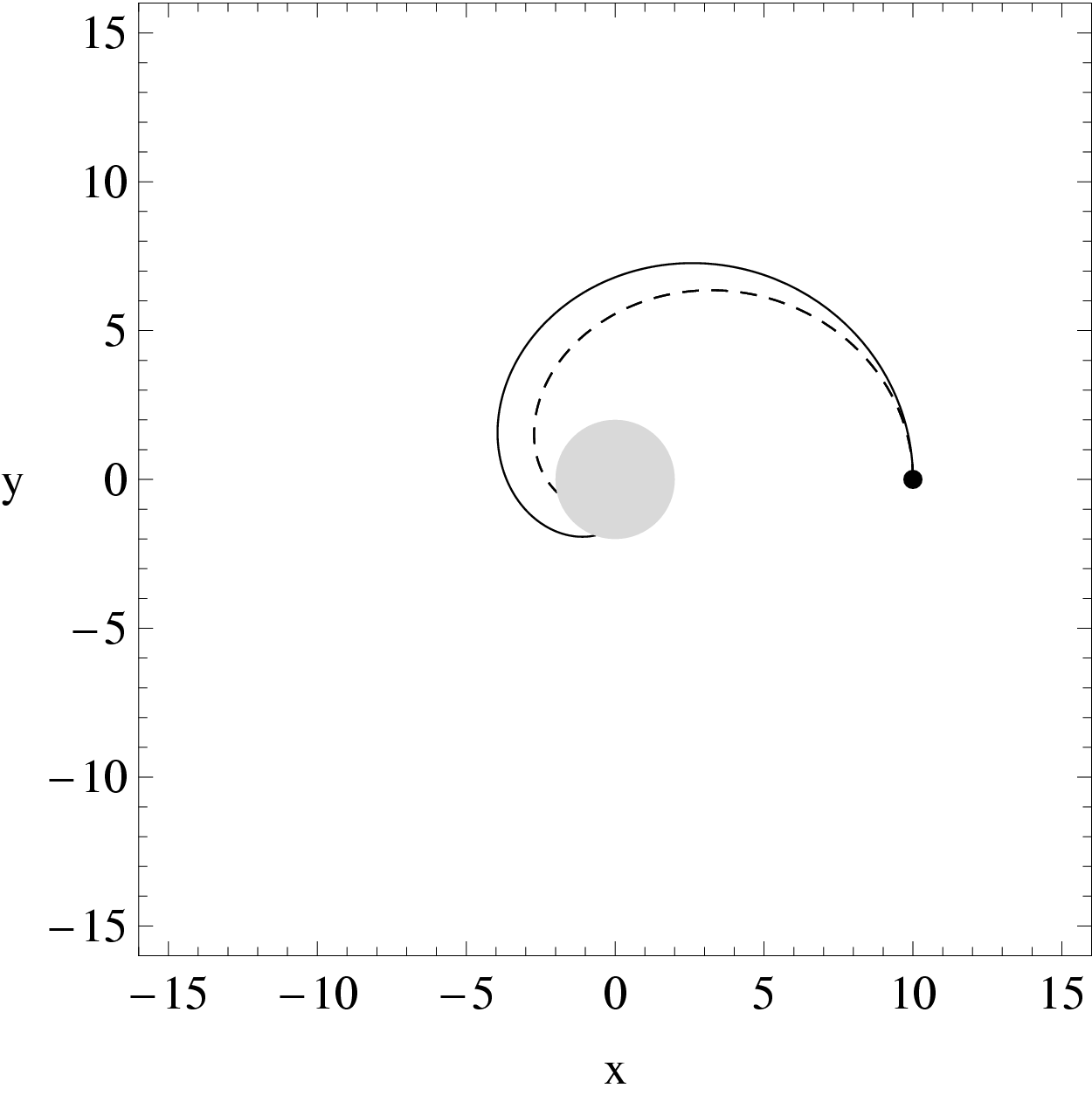}} \subfigure[
\quad BO $\LL=4$]{\includegraphics[width=0.3\hsize]{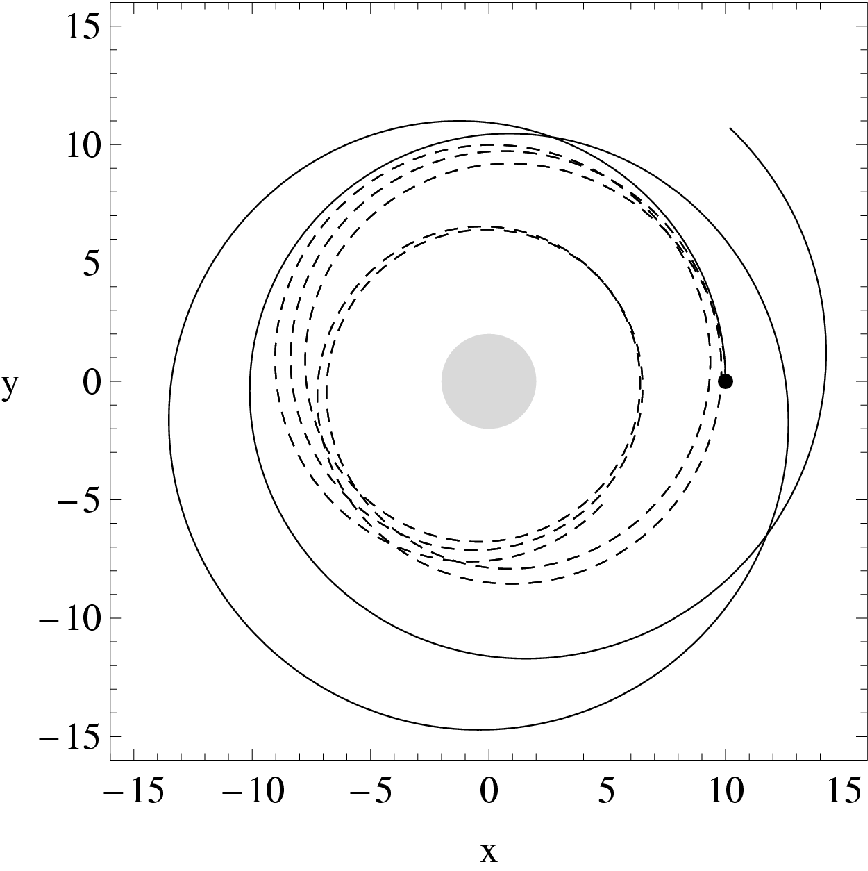}}
\subfigure[ \quad EO
$\LL=6$]{\includegraphics[width=0.3\hsize]{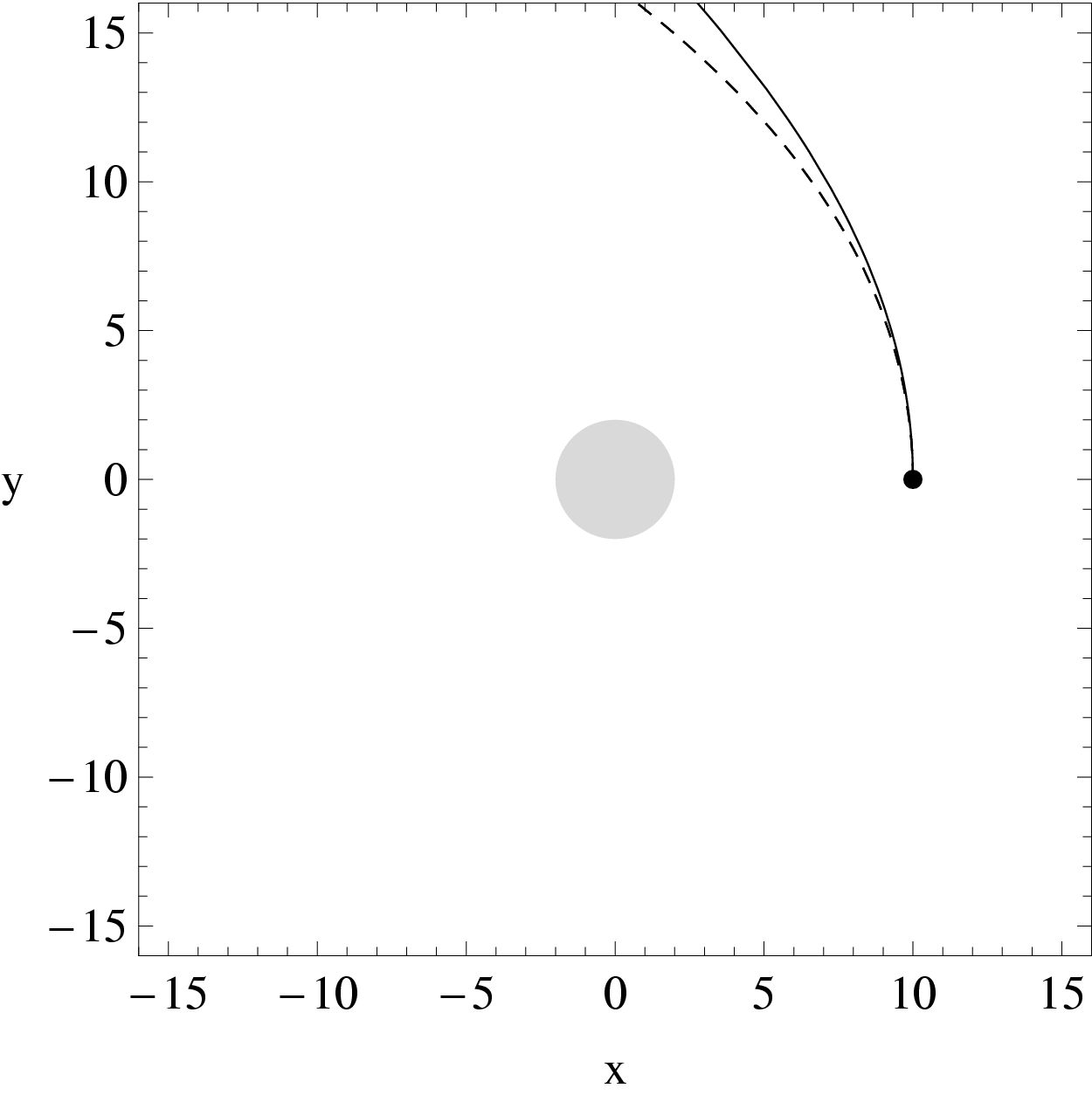}}
\caption{Examples of particle trajectories moving at the
equatorial plane around black string. We compare motion with
${\cal J}=0$ (solid curve) with ${\cal J}=0.5$ (dashed) for all
TO, BO and EO possible orbits. Particles start from the position
$r_0=10$ but with the different energies. \label{figNO1}}
\end{figure*}

The type of the motion depends on the number of the turning points
i.e. on the number of extrema of the effective potential. The
extremum of the effective potential lies at
\begin{equation}
r_{\rm ext} = \frac{{\cal L}^2 \pm {\cal L} \sqrt{{\cal L}^2
-12\left(1+{\cal J}^2\right)}}{2\left(1+{\cal J}^2\right)}.
\label{rextrST}
\end{equation}
Hereafter we set the mass density of the black string as $M=1$. It
is obvious from (\ref{rextrST}) that the condition of the presence
of the extrema of the effective potential corresponds to the
positive values of the radicand of (\ref{rextrST}), which gives
\begin{equation}
|{\cal L}| > 2\sqrt{3}\sqrt{1+{\cal J}^2}.
\end{equation}
Numbers of extrema define the possibility of the existence of
turning points. The samples of the trajectories are shown in
Fig.~\ref{figNO1} for the different values of {parameter} ${\cal
J}$.

\subsection{Freely falling particle} \label{freeparSBS}

\begin{figure*}
    \includegraphics[width=5.5cm]{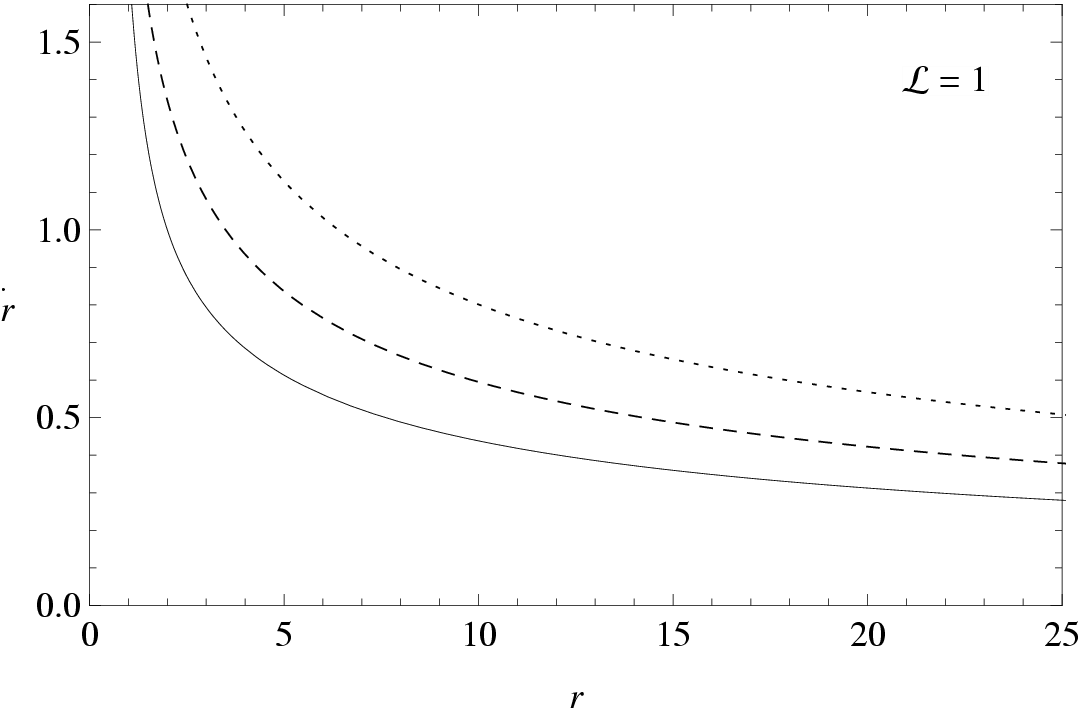}  \label{pic:acceleration-a}
    \includegraphics[width=5.5cm]{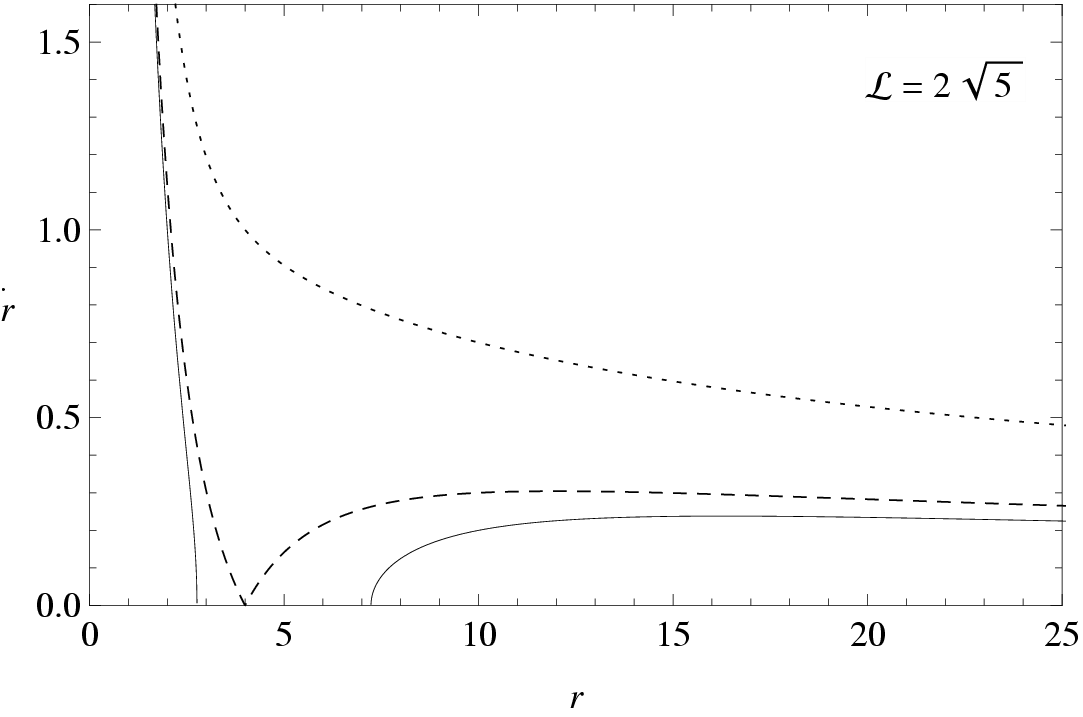}  \label{pic:acceleration-b}
    \includegraphics[width=5.5cm]{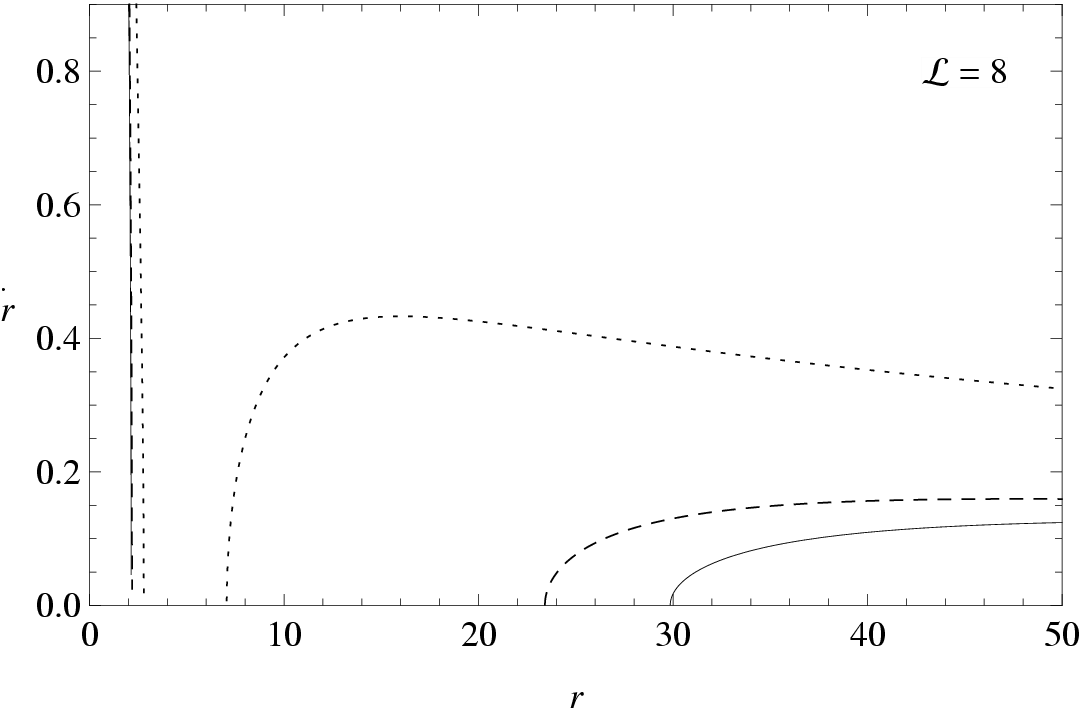}  \label{pic:acceleration-c}
  \caption{Acceleration of a freely falling particle along the axis $r$ by the static black string
  for three different values of parameter $\JJ$: $\JJ=0$ (solid), $\JJ=0.5$ (dashed) and $\JJ=1.5$ (dotted) and three representative values
 of the angular momentum: left panel corresponds
 to the case when $\LL=1$, middle panel corresponds to the case when $\LL=4.47$ and right panel corresponds to the case when $\LL=8$.
 The critical case corresponds to the dashed curve in the middle panel.}
 \label{pic:acceleration}
\end{figure*}

Consider a neutral particle freely falling {from infinity at rest}
in the direction of an event horizon of the static black string.
The final state of the particle entirely depends on its momenta
$\LL$ and $\JJ$. If the angular momentum of the particle $\LL$ is
larger than {maximum} critical value, then the geodesics of the
particle may never reach the vicinity of the horizon and the
strong gravitational {effects} cannot be analyzed. On the other
hand, for the small values of angular momentum, the particles fall
radially with a small tangential velocity and the center-of-mass
energy does not grow either. Consequently, there is a critical
value for the angular momentum such that particles may reach the
horizon with maximum tangential velocity. The energy of the
particle at rest at infinity corresponds to its energy in flat
spacetime
\begin{equation}
\EE^2=1+\JJ^2, \label{enerflat}
\end{equation}
which is in contrast with the value $\EE_{\rm (4D)} = 1$ {defined}
in four-dimensional spacetime . To find the value of the critical
angular momentum one can rewrite Eq.~(\ref{eqn:r-equation}) using
(\ref{enerflat}) in the form
\begin{equation}
\frac{dr}{d\tau} = \dot{r} = \pm \frac{1}{r^2} \sqrt{2r^3
\left(1+\JJ^2\right) - \LL^2 r(r-2)}.
\end{equation}
The dependence of $\dot{r}$ on the radial coordinate $r$ for
different values of $\LL$ and $\JJ$ is shown in
Fig.\ref{pic:acceleration}. It is obvious from
Fig.~\ref{pic:acceleration} that there is a critical value of the
angular momentum
\begin{equation}
\LL_{\rm cr} = \pm 4\sqrt{1+\JJ^2}\ , \label{LcritSTAT}
\end{equation}
{for} which the particle still may reach the horizon of black
string.

In the case of the Schwarzschild spacetime ($\JJ=0$) the critical
angular momentum reduces to the {standard} value
\begin{equation}
\LL_{\rm cr (4D)} = \pm 4. \label{Lpm4D}
\end{equation}

One may see from Fig.~\ref{pic:potential} the increase of critical
angular momentum of particles occurs in the presence of
nonvanishing  parameter $\JJ$. The graph corresponding to the
critical angular momentum, which is indicated with the thick line,
shifts upwards with the increase of {parameter} $\JJ$. The
presence of parameter $\JJ$ causes the particle orbits to become
more unstable compared to the case $\JJ=0$. Bound and escape
orbits are forced to become terminated orbits due to the presence
of nonvanishing parameter $\JJ$. This effect is shown in more
clear way in Fig.~\ref{figNO1}. The dashed line corresponding to
the case $\JJ=0.5$ is more unstable compared to the case when
$\JJ=0$.

\subsection{Collision of freely falling particles}

In this section we {study} the collision of two particles near the
horizon of the nonrotating black string. Consider two particles
approaching the black string with the different angular momenta
${\cal L}_1$, ${\cal J}_1$ and ${\cal L}_2$, ${\cal J}_2$ and
colliding at some radius $r$ at the equatorial plane of the black
string. {Assume} that the particles are at rest at infinity; i.e.
their energies satisfy Eq.~(\ref{enerflat}). In the center-of-mass
frame the maximal energy will have a place where the particles
collide near a black string with maximal and opposite angular
momenta given by the expression (\ref{LcritSTAT}).

Our aim is to compute the energy in the center-of-mass frame for
the collision of two particles. Since the background is curved, we
need to define the center-of-mass frame properly. It turns out
that there is a simple formula for $E_{\rm c.m.}$ valid both in
flat and curved spacetimes:
\begin{equation}
E_{\rm c.m.} = m_0 \sqrt{2}\sqrt{1-g_{\alpha\beta} U^\alpha_{(1)}
U^\beta_{(2)}}, \label{encolstat}
\end{equation}
where $U^\alpha_{(1)}$ and $U^\beta_{(2)}$ are the five-velocities
of the particles, properly normalized by $g_{\alpha\beta} U^\alpha
U^\beta = -1$. This formula is of course well known in special
relativity, and the principle of equivalence should be enough to
ensure its validity in a curved background \cite{BSW:2009}.
Components of $U^\alpha_{(1)}$ and $U^\beta_{(2)}$ can be
straightforwardly calculated from (\ref{eqn:t-equation}) -
(\ref{eqn:r-equation}) in the form

\begin{eqnarray}
U^\alpha_{(1)} = \hspace{7.77cm} \\
\Bigg( \frac{(1+\JJ_1^2) r}{r-2},~\pm\frac{\sqrt{2r^3(1+\JJ_1^2)-\LL_1^2
r(r-2)}}{r^2},~0,~\frac{\LL_1}{r^2},~\JJ_1 \Bigg),\nonumber
\\
 U^\alpha_{(2)} = \hspace{7.77cm} \\
\Bigg( \frac{(1+\JJ_2^2)
r}{r-2},~\pm\frac{\sqrt{2r^3(1+\JJ_2^2)-\LL_2^2
r(r-2)}}{r^2},~0,~\frac{\LL_2}{r^2},~\JJ_2 \Bigg),\nonumber
\end{eqnarray}
{where the sign $"+"$ corresponds to incoming particles, while the
sign $"-"$ corresponds to outgoing ones.} Here we take the energy
of the particle at rest at infinity, i.e. $\EE^2=1+\JJ^2$. Thus
the energy of two colliding particles with the same mass $m$ near
static black string in the center-of-mass frame is given by
\begin{eqnarray}
\frac{1}{2m^2} E_{\rm c.m.}^2& = &1 - \JJ_1 \JJ_2-\frac{\LL_1
\LL_1}{r^2}+ \frac{r \sqrt{1+\JJ_1^2}\sqrt{1+\JJ_2^2}}{r-2} {}
 \nonumber\\
&&-\frac{\sqrt{2r^2(1+\JJ_1^2)-\LL_1(r-2)}
}{r^2(r-2)} \nonumber\\&&\times\sqrt{2r^2(1+\JJ_2^2)-\LL_2(r-2)}. \label{ecmschw}
\end{eqnarray}

\begin{figure*}[t!]
 \includegraphics[width=7cm]{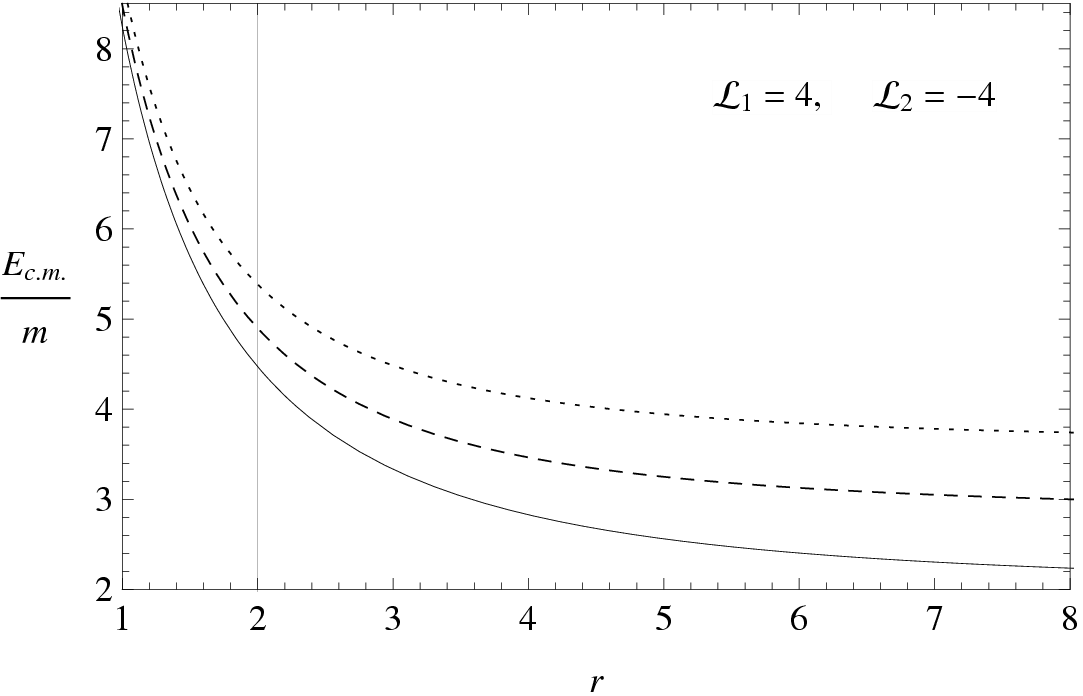}
   \includegraphics[width=7cm]{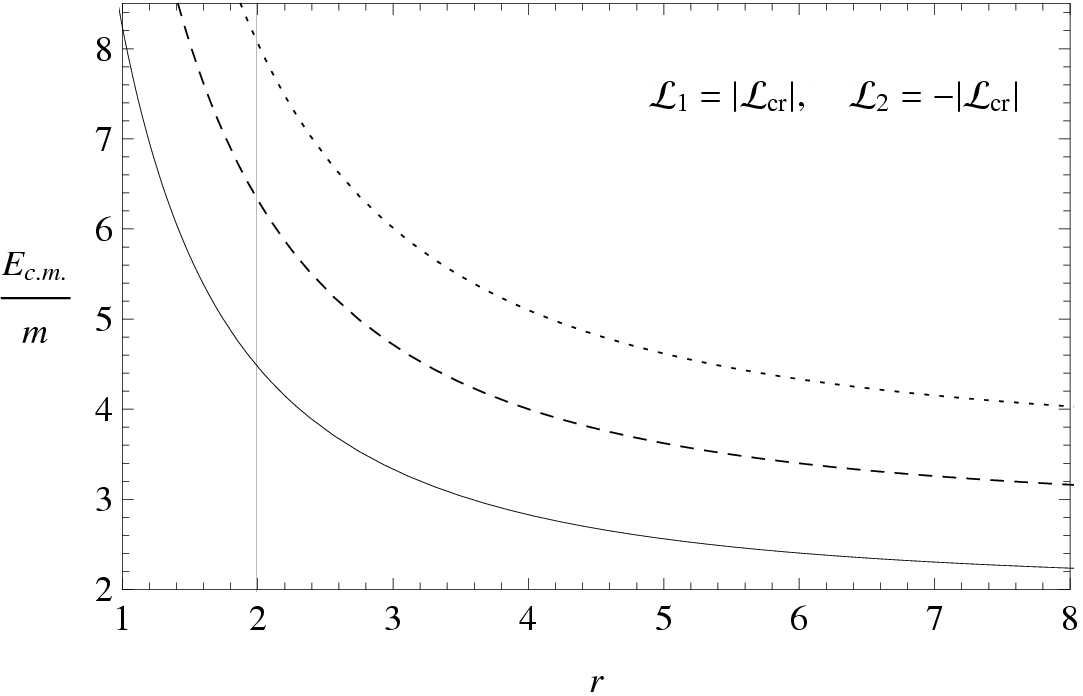}
 \caption{The variation of the energy of the collision of two particles with the opposite
 momenta near the static black string in the center-of-mass frame for three different values
 of parameter $\JJ$. $\JJ_{1,2}=0$ (solid), $\JJ_{1,2}=\pm 1$ (dashed),
  $\JJ_{1,2}=\pm 1.5$ (dotted). The event horizon is indicated by the vertical thin line. }
 \label{pic:ecm_stat_bs}
\end{figure*}

Ex facte, one can say that Eq.~(\ref{ecmschw}) diverges at the
singularity $r=0$, the horizon of black string is given by $r=2$
and unlimited collisional energy appears in the case of the
collision at the horizon. However, it is not satisfied (see
Fig.~\ref{pic:ecm_stat_bs}) since the numerator of (\ref{ecmschw})
also vanishes at this point. If we observe near the horizon
{where} $r \rightarrow 2$, collision of two particles falling on
the black string with critical angular momenta given by
(\ref{LcritSTAT}), the energy of the collision will {be reduced}
to the value
\begin{eqnarray}
E_{\rm c.m.} = m \sqrt{2} \sqrt{1-\JJ_1 \JJ_2 + 9
\sqrt{\left(1+\JJ_1^2\right)\left(1+\JJ_2^2\right)}}\ , \ \
\label{ecmRZschw}
\end{eqnarray}
where $m$ is the mass of the colliding particles. Evidently the
maximal energy of the collision occurs if the values of $\JJ_1$
and $\JJ_2$ are opposite. In the special symmetric case when the
parameters $\JJ_{1,2}$ have the same and opposite values:
$\JJ_{1}=-\JJ_{2}=\JJ$, the expression (\ref{ecmRZschw}) for the
maximal collision energy per unit mass $(\EE_{\rm c.m.} = E_{\rm
c.m.}/m)$ reduces to the following simple form {
\begin{equation} \EE_{\rm c.m.}^{\rm max} = 4.47
\sqrt{1+\JJ^2}. \label{ecmRZschwMAX}
\end{equation}
In the absence of the additional dimension, i.e. when $\JJ=0$, the
maximal energy of the collision in the center-of-mass frame
reduces to $\EE_{\rm c.m.}^{\rm max} = 2\sqrt{5}$, which coincides
with {the value defined in} the Schwarzschild black hole spacetime
\cite{BSW:2009}. }

\begin{figure}[h]
 \includegraphics[width=7cm]{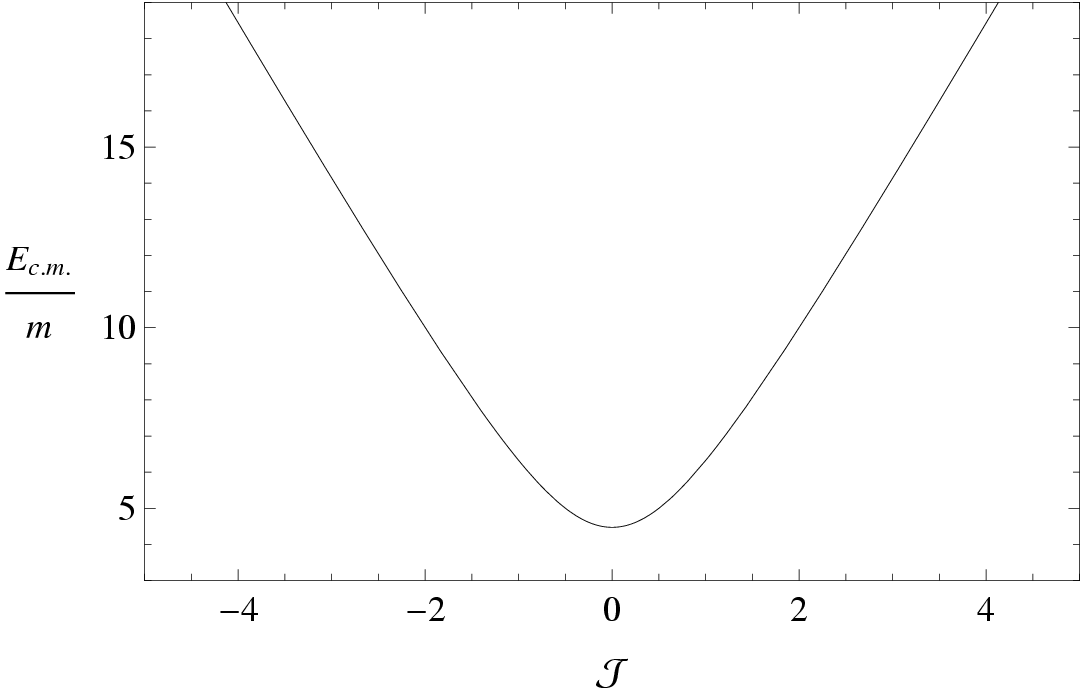}

 \caption{Maximal energy of the collision as function
 of the parameter $\JJ$.}
 \label{Ecm_sch_as_J}
\end{figure}

It is easy to see from the Fig.\ref{Ecm_sch_as_J} that the
collisional energy infinitely grows when the parameter $\JJ$
tends to infinity and formally, one can obtain an arbitrary large
energy in such collisions of particles. However, for our purposes
this case is not interesting, since the energy of such particles
measured at infinity also grows in this limit and the high
collision energy is possible only if the
initial energy of the particles measured at infinity is also extremely high.

\section{Black string immersed in magnetic field as accelerator of charged particles\label{secIII}}

\subsection{Dynamical equations}

Here we study a charged particle motion in the vicinity of a black
string of mass {density} $M$ given by the spacetime metric
(\ref{statmetric}) in the presence of an external static
axisymmetric and asymptotically uniform magnetic field.

Since the spacetime metric (\ref{statmetric}) is Ricci flat the
commuting Killing vectors $\xi_{(t)}$ and $\xi_{(\varphi)}$
{satisfy} the equations
\begin{equation}
\square \xi^\alpha = 0\ .
\end{equation}
These equations coincide with the Maxwell equations $\square
A^\alpha = 0$ for a vector potential $A^{\alpha}$ of the
electromagnetic field in the Lorentz gauge. In what follows, we
use the simplest {but realistic} case {when} the magnetic field is
considered as a test field in the given gravitational background
of the nonrotating black string, in such sense that the magnetic
field is homogeneous at the spatial infinity where it has the
strength $B$ \cite{Frolov:2010,wald}. For such configuration of
the magnetic field one can easily find the components of the
vector potential of the electromagnetic field in the form
\begin{equation}
A^{\alpha} = \frac{B}{2} \xi^{\alpha}_{(\varphi)}.
\end{equation}
Keeping the symmetry hereafter, we assume that $B\geq0$. {Finally
the five-vector potential $A_{\alpha}$ of the electromagnetic
field will take a form
\begin{equation}
A_t = A_r = A_{\vartheta} = A_w = 0, \quad A_{\varphi} =
\frac{1}{2} B r^2 \sin^2{\vartheta}.
\end{equation}
}

 Associated with a
timelike $\xi^{\mu}_{(t)}$ and spacelike $\xi^{\mu}_{(\varphi)}$
and $\xi^{\mu}_{(w)}$ Killing vectors one can find the following
conserved quantities along a geodesics as
\begin{eqnarray}
\EE& =& \frac{E}{m} = -\xi^{\mu}_{(t)} \frac{P_{\mu}}{m} =
\left(1-\frac{2M}{r}\right) \frac{dt}{d\tau}, \label{MF-intE}\\
\LL &=&\frac{L}{m} = \xi^{\mu}_{(\varphi)} \frac{P_{\mu}}{m} = r^2
\sin^2{\vartheta} \left( \frac{d\varphi}{d\tau}  + \frac{q B}{2
m}\right), \label{MF-intL}\\
\JJ &= &\frac{J}{m} = \xi^{\mu}_{(w)} \frac{P_{\mu}}{m} = \frac{d
w}{d\tau}. \label{intJ}
\end{eqnarray}
Here $P_{\mu} = m u_{\mu}+q A_{\mu}$ is the generalized
five-momentum of the particle with mass $m$ and charge $q$. {In
order to study the charged particle motion in combined
gravitational and magnetic fields one can use the Hamilton-Jacobi
equation in the following form
\begin{equation}
g^{\mu\nu} \left(\frac{\partial S}{\partial x^{\mu}} + q A_{\mu}
\right) \left(\frac{\partial S}{\partial x^{\nu}} + q A_{\nu}
\right) = -m^2, \label{Ham-Jac-Mag}
\end{equation}
where $q$ and $m$ are the charge and mass of the particle,
respectively. {The Hamiltonian of the charged particle in a
magnetic field has the following form
\begin{equation}
2 H = g^{\alpha\beta} (p_{\alpha} + q A_{\alpha}) (p_{\beta} + q
A_{\beta}) + m^2,
\end{equation}
where $p_{\mu} = m u_{\mu}$ is the five momentum of the particle.}

Since $t$, $\varphi$ and $w$ are the Killing variables one can
write the action in the form
\begin{equation}
S = - \EE t + \LL \varphi +\JJ w + S_{r\vartheta} (r, \vartheta),
\label{Action-Mag}
\end{equation}
where the conserved quantities $\EE$, $\LL$ and $\JJ$ are the
energy, angular momentum and new momentum of a test particle
measured at infinity which is related to the additional dimension.
Substituting (\ref{Action-Mag}) into Eq.~(\ref{Ham-Jac-Mag}) one
may get the equation for inseparable part of the action in the
form
\begin{eqnarray}
&&-\left(1-\frac{2M}{r}\right)^{-1}
\EE^2+\left(1-\frac{2M}{r}\right)\left(\frac{\partial
S_{r\vartheta}}{\partial r}\right)^2
+\frac{1}{r^2}\left(\frac{\partial S_{r\vartheta}}{\partial\vartheta}\right)^2{}
  \nonumber \\
&&+\frac{1}{r^2 \sin^2{\vartheta}} \left(\LL + \frac{q B}{2} r^2
\sin^2{\vartheta}\right)^2 + \JJ^2+m^2 = 0.
\end{eqnarray}
It is quite easy to separate variables in this equation at the
equatorial plane $\vartheta=\pi/2$ and obtain the differential
equations for each coordinate in the following form}
\begin{eqnarray}
\frac{dt}{d\tau}& =& \frac{\EE r}{r-2M}, \label{mag-t-eq}
\\
\frac{d\varphi}{d\tau} & =& \frac{\LL}{r^2} - \BB, \label{mag-phi-eq}
\\
\frac{d w}{d \tau} & =& \JJ, \label{mag-w-eq}
\\
\left(\frac{d r}{d \tau}\right)^2 & =& \EE^2 - V_{\rm eff},
\label{mag-r-eq}
\end{eqnarray}
where the effective potential has a form
\begin{equation}
V_{\rm eff} = \left(1-\frac{2M}{r}\right)
\left[1+\JJ^2+\left(\frac{\LL}{r} - \BB r\right)^2\right].
\label{mag-veff}
\end{equation}
{Here we denote new parameter characterizing the coupling between
the external magnetic field and the test charged particle as}
\begin{equation}
\BB = \frac{qB}{2m}.
\end{equation}
 Depending
on {the direction of the Lorentz} force acting on the charged
particle the angular momentum of the particle $\LL$ can be either
positive or negative. The direction of the {Lorentz} force does
not depend on the parameter $\JJ$, which may, however, change its
absolute value. When $\LL > 0$, the Lorentz force acting on a
charged particle is repulsive; i.e., it is directed outward from
the black string, and when $\LL < 0$ the Lorentz force is
attractive i.e. it is directed {towards} the black string.

We study the characteristic properties of the motion of charged
particles in the exterior spacetime of magnetized black string
when $r>2M$. Expression (\ref{mag-veff}) shows that the effective
potential is positive in that region and vanishes at the horizon
of the black string where $r = 2 M$. But when $r \rightarrow
\infty$, the effective potential grows as $\BB^2 r^2$. This
property implies that the particle never reaches the spatial
infinity and its motion is always finite. {Furthermore}, the
{Lorentz} force acting on the charged particle forces the particle
to revolve around the axis of the external magnetic field. Due to
this it is very important to study the circular motion of the
charged particle around black string and especially for the
question of the existence of the innermost stable circular orbits.

\subsection{Circular motion of a charged particle}

The minimum of the effective potential corresponds to the circular
motion of a particle. Consider a circular motion of a charged
particle around a black string. Taking the mass {density} of the
black string as $M=1$, one can write the momentum of a particle at
the circular orbit of radius $r$ as
\begin{equation}
p^\mu = m \gamma \left(\sqrt{\frac{r}{r-2}} \delta^{\mu}_{(t)} +
\frac{v}{r} \delta^{\mu}_{(\varphi)} + \delta^{\mu}_{(w)}\right).
\label{pmucharged}
\end{equation}
Here $v$ (which can be both positive and negative) is a velocity
of the particle with respect to a rest frame and
\begin{equation}
\gamma = \frac{1}{\sqrt{1-v^2}}
\end{equation}
is the Lorentz gamma factor.

Using the expression $d\varphi/d\tau = v\gamma/r$ together with
(\ref{MF-intL}) and (\ref{mag-phi-eq}) one can find
\begin{equation}
v\gamma = \frac{\LL-\BB r^2}{r}\ ,
\end{equation}
and easily obtain the expressions for the velocity and the
{Lorentz} gamma factor as
\begin{equation}
v = \frac{\LL-\BB r^2}{\sqrt{r^2+\left(\LL-\BB r^2\right)^2}},
\end{equation}
\begin{equation}
\gamma^2 = 1 + \left(\frac{\LL}{r} - \BB r \right)^2.
\label{MF-gamma-L-B-r}
\end{equation}
%

\begin{figure*}[t!]
 \includegraphics[width=7cm]{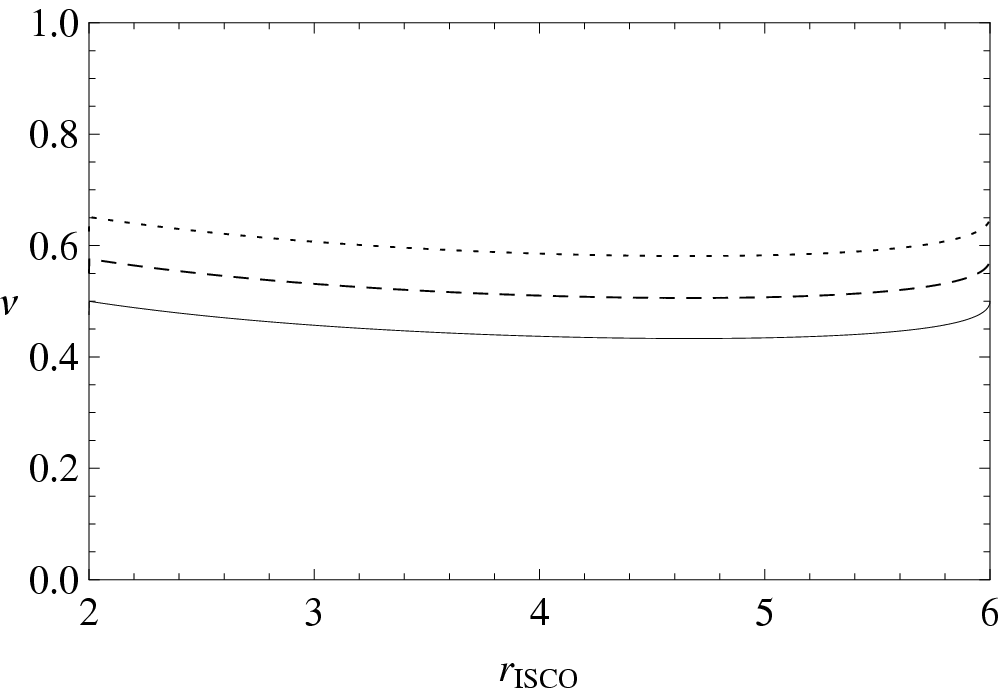}  \label{mf-velgam-vel}
 \includegraphics[width=7cm]{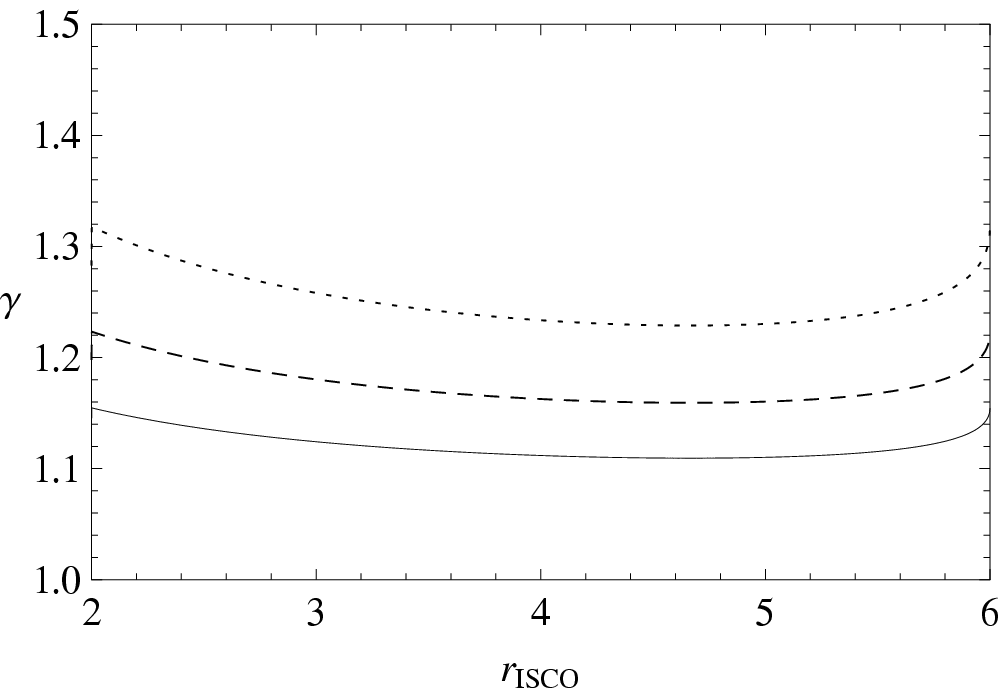}  \label{mf-velgam-gam}
  \caption{Velocity of the particle and its Lorenz gamma factor at the ISCO as the function
 of its radius for the positive angular momentum and different values of parameter $\JJ$:
 $\JJ=0$ (solid), $\JJ=0.7$ (dashed) and $\JJ=1.1$ (dotted).}
 \label{mf-velgam}
\end{figure*}

\begin{figure}[h]
 \includegraphics[width=7cm]{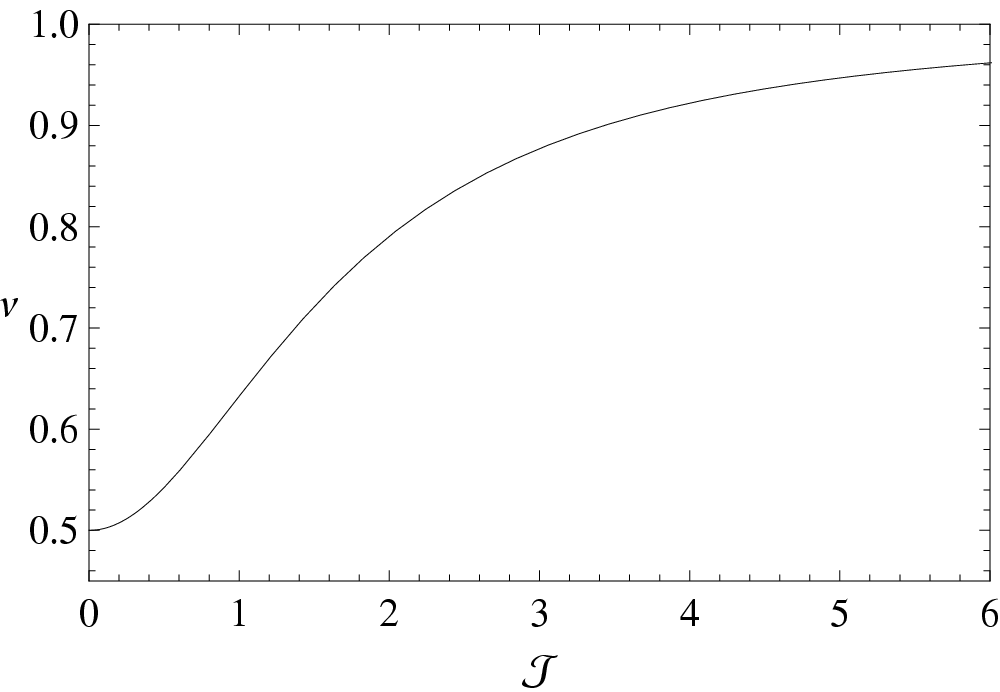}

 \caption{Velocity of the particle at the ISCO as the function
 of parameter $\JJ$.}
 \label{vel-J-isco6}
\end{figure}

Plots of the velocity and the {Lorentz} gamma factor at the
innermost stable circular orbits (ISCOs) are shown in
Fig.~\ref{mf-velgam} for the different values of the parameter
$\JJ$. It is obvious from Figs.~\ref{mf-velgam} and
\ref{vel-J-isco6} that the parameter $\JJ$ increases the velocity
of the particle and the $\gamma$ factor is maximal at the end
points of the domain of the ISCO radius. The domain of the
definition of the ISCO radius is restricted by the region
$2<r_{\rm ISCO}<6$ (see the left panel of Fig.~\ref{rIsco_B}).
Consequently, the limiting value of the particle velocity at the
end of the ISCO domain is
\begin{equation}
v = \sqrt{\frac{1 + J^2}{4 + J^2}},
\end{equation}
which tends to the speed of light ($v \rightarrow 1 = c$) when
$\JJ \rightarrow \infty$. Dependence of the velocity of the
particle at ISCO on the parameter $\JJ$ is presented in
Fig.~\ref{vel-J-isco6}. Similarly, the gamma factor of the
particle has the maximal value at the end points of the domain of
ISCO, namely, at $r=2$ and $r=6$. At these positions the gamma
factor takes the value
\begin{equation}
\gamma_{\rm max} = \frac{1}{\sqrt{3}}\sqrt{4 + J^2},
\label{mf-gamMAX}
\end{equation}
which tends to infinity if $\JJ \rightarrow \infty$.

\subsection{Innermost stable circular orbits}

The ISCO is defined by the equations when the first and second
derivatives of the effective potential are equal to zero:
\begin{equation}
\BB^2 r^4 (r-1) - 2 \BB \LL r^2 + \left(1+\JJ^2\right) r^2 - \LL^2
r + 3 \LL^2 = 0, \label{Dveff}
\end{equation}
\begin{equation}
\BB^2 r^5 + 4 \BB \LL r^2 - 2 \left(1+\JJ^2\right) r^2 + 3 \LL^2 r
- 12 \LL^2 = 0. \label{DDveff}
\end{equation}
These equations allow one to find the location of the ISCO and
parameter $\LL$ {for given values of} $\BB$ and $\JJ$. Summation
of (\ref{Dveff}) and (\ref{DDveff}) {gives} the expression for the
angular momentum in terms of $\BB$ and $r$ as
\begin{equation}
\LL = \pm \BB r^2 \frac{\sqrt{3 r-2}}{\sqrt{6-r}}.
\label{LL-ISCO-BB}
\end{equation}
Substituting this equation into (\ref{DDveff}) one can find the
interdependency of the magnetic field and ISCO radius for the
fixed $\JJ$ as follows
\begin{equation}
\BB =  \frac{\sqrt{(1 + J^2) (6 - r)}}{r \sqrt{2} \sqrt{ 2 r^2 - 9
r +6 \pm \sqrt{(6 - r)(3 r - 2)}}} \ . \label{BB-ISCO-JJ}
\end{equation}
The condition that $\BB$ is real imposes the restrictions on the
possible values of the ISCO radius:
\begin{eqnarray}
&2 < r_{\rm ISCO} \leq 6, \qquad &\mbox{for} \quad \LL>0,
\\
&\frac{1}{2} \left(5 + \sqrt{13}\right) < r_{\rm ISCO} \leq 6,
\qquad &\mbox{for} \quad \LL<0.
\end{eqnarray}
%

\begin{figure*}[t!]
   \includegraphics[width=7cm]{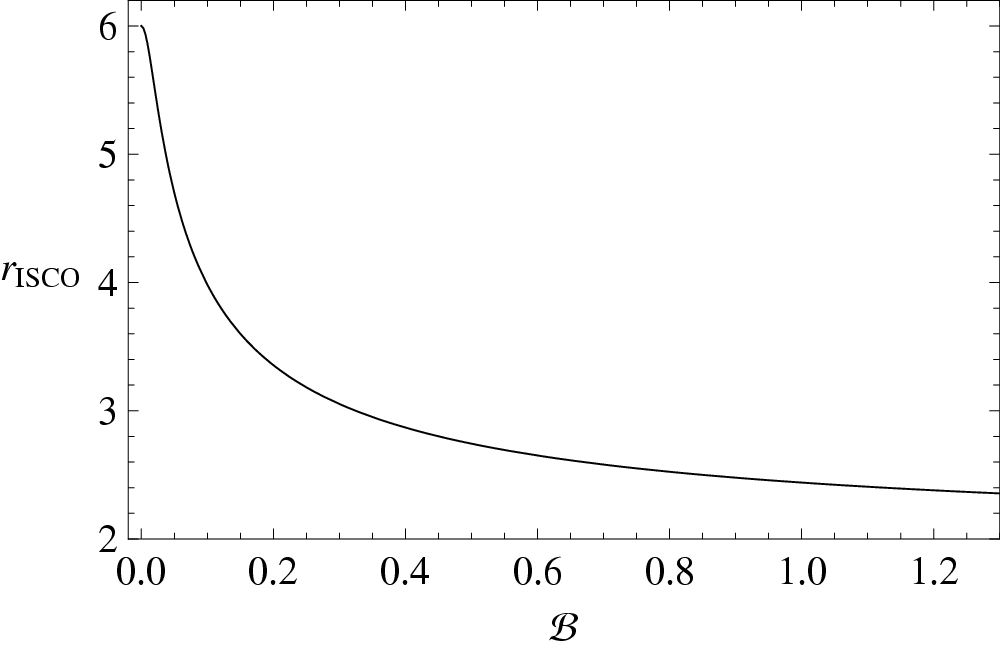}
   \includegraphics[width=7cm]{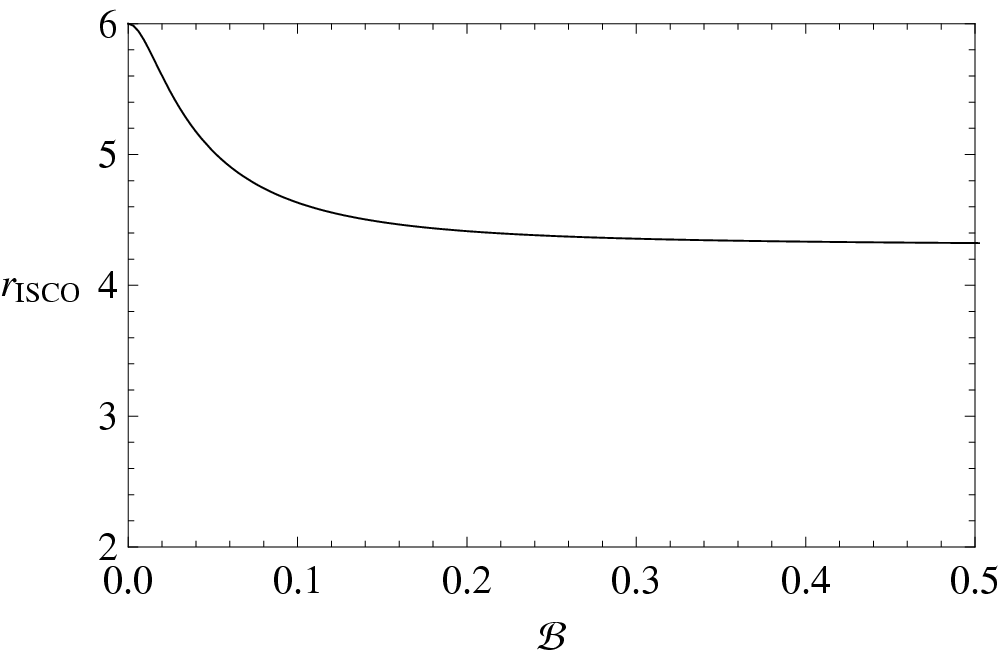}
 \caption{Dependence of the innermost stable circular orbits
 $r_{\rm ISCO}$ on the magnetic field parameter $\BB$. The left panel
 corresponds to the motion of the particle with positive angular
 momentum while right panel corresponds to the particle with negative angular
 momentum $\LL$.}
 \label{rIsco_B}
\end{figure*}

The dependence of the ISCO radii on the magnetic field strength is
shown in Fig.\ref{rIsco_B} for two different cases of particle
motion: $\LL>0$ and $\LL<0$, when the influence of the Lorentz
force on the particle is opposite. One can see from
Fig.\ref{rIsco_B}, that the magnetic field shifts the ISCO radius
towards the horizon of black string. In the limit of the strong
magnetic field, i.e., when $\BB \gg 1$, one can find the value of
the ISCO radius for the positive angular momentum as
\begin{equation}
r_{\rm ISCO} \approx r_{\rm H} + \frac{\sqrt{1 + J^2}}{\sqrt{3}
\BB}, \qquad \BB \gg 1 \ ,\label{risco-approx}
\end{equation}
which means that the particle is moving in strong magnetic field
and the repulsive Lorentz force has the radius of the innermost
stable orbit very close to the horizon of the black string.
{Recalling} that $\BB=q B/2m$, one can conclude that the ISCO for
electrons is located much closer to the horizon than the
corresponding orbits for protons. Under the influence of the
parameter $\JJ$, the minimal value of the radius of circular
orbits increases and charged particles may be shifted towards an
observer at infinity which means that the orbits of the charged
particles may become unstable.

\subsection{Collision of particles}

In this section, we deviate somewhat from the original idea of BSW
\cite{BSW:2009}. We consider two different situations. In the
first case a charged particle orbiting at the ISCO collides with
another charged particle at the ISCO. In the second case, we
consider the collision of a freely falling from infinity neutral
particle with a charged particle revolving at the ISCO instead of
a near-horizon collision.

\subsubsection{Head-on collision of charged
particles orbiting at ISCO}

Hereafter, we use {the approach} and methodology properly
developed by Frolov~\cite{Frolov:2012}. As a first example we
consider a head-on collision of two identical charged particles
moving at the same circular orbit in opposite directions with the
same mass $m$, opposite charges $+|q|$ and $-|q|$ and the opposite
{signs of the parameter} $\JJ$ and $-\JJ$. The four-momentum of
the system after the collision is $P^{\mu} = 2 m \gamma
(1-2/r)^{-1/2} \xi^{\mu}_{(t)}$. Then the energy after the
collision calculated in the center-of-mass frame $E_{\rm c.m.}$
has the following form \cite{Frolov:2012}:
\begin{equation}
E_{\rm c.m.} = 2 m \gamma. \label{mf-ecm2mg}
\end{equation}
For the particle orbiting at the ISCO around a static black string
with the fixed {parameter} $\JJ$, the gamma factor remains almost
unchanged along the range of the ISCO for any value of the
strength of the magnetic field (see Fig.~\ref{mf-velgam}). A
substitution of the maximal value of the gamma factor given by
Eq.~(\ref{mf-gamMAX}) into (\ref{mf-ecm2mg}) gives the equation
for the maximal center-or-mass energy per unit mass $(\EE_{\rm
c.m.} = E_{\rm c.m.}/m)$ in the form
\begin{equation}
\EE_{\rm c.m.}^{\max} = 1.15 \sqrt{4+\JJ^2}. \label{mf-ecm2mgMAX}
\end{equation}
For the minor values of the parameter $\JJ$, the energy of the
collision is slightly higher than $2 m$. However, similarly to the
case of the absence of the magnetic field given by
Eq.(\ref{ecmRZschwMAX}) it is easy to see that the energy of the
collision of two particles initially revolving at the ISCO
infinitely grows when the parameter $\JJ$  tends to infinity and
formally one can obtain an arbitrary large energy. {However}, it
does not provide the colliding particles with high energy since
the energy of such particles as measured at infinity also grows in
this limit.

\subsubsection{Collision of a freely falling
neutral particle with a charged particle at ISCO}

Now we consider the collision of two particles when one of them is
charged and revolved at the ISCO and the other one is neutral and
freely falls from infinity in the direction of the black string
horizon. We assume that both particles move at the equatorial
plane $\vartheta=\pi/2$ of the black string. In the center-of-mass
frame the maximal energy would have place when the particles
collide near a black string with maximal and opposite angular
momenta. For a neutral particle this condition {is defined by the
expression} (\ref{LcritSTAT}). At the moment of the collision, the
momentum is the sum
\begin{equation}
P^{\alpha} = p_{\rm c}^{\alpha} + p_{\rm n}^{\alpha},
\end{equation}
where $p_{c}^{\alpha}$ and $p_{n}^{\alpha}$ are momenta of charged
and neutral particles, respectively. This corresponds to the
collisional center-of-mass energy which can be written in the form
\begin{equation}
E_{\rm c.m.}^2 = m_c^2 + m_n^2 - 2g_{\mu\nu} p_c^{\mu} p_n^{\nu}.
\end{equation}
Using the equations (\ref{eqn:t-equation})-(\ref{eqn:r-equation})
and (\ref{pmucharged}) one can find the center-of-mass energy of
the collision as
\begin{equation}
E_{\rm c.m.}^2 = m_c^2 + m_n^2 + 2 m_c \gamma_c  \left( \EE_n
\sqrt{ \frac{r}{r-2} } - \frac{v_c \LL_n}{r} - \JJ_{n}\right),
\label{ecmdecomposed}
\end{equation}
where subscript $n(c)$ is responsible for neutral(charged)
particles. Since the radius of the ISCO can be arbitrarily close
to the horizon, the first term in the brackets can be made
arbitrarily large while $\gamma$ remains finite (see
Fig.\ref{mf-velgam}~(b)). The second and the third terms in the
brackets are finite for the typical values of {the parameter}
$\JJ$ (see Eq.(\ref{LcritSTAT}) and Fig.\ref{mf-velgam}~(a)).
{Consequently}, the leading contribution to the energy of the
collision near the horizon can be written in the following form:
\begin{equation}
E_{\rm c.m.}^2 \approx \frac{2\sqrt{2} m_c \gamma_c
\EE_n}{\sqrt{r-2}} \qquad (r\rightarrow2). \label{ecmmaincontr}
\end{equation}
Using the relation (\ref{risco-approx}), one can obtain the
asymptotic value of the center-of-mass energy for a collision of a
neutral particle with the charged one at ISCO in a magnetic field
$\BB \gg 1$ as
\begin{equation}
E_{\rm c.m.}^2 \approx 2\sqrt{2} m_c \gamma_c \EE_n
\left(\frac{\sqrt{3} \BB}{\sqrt{1+\JJ^2}}\right)^{1/2}.
\label{ecmbigBB}
\end{equation}
Using Eqs. (\ref{MF-gamma-L-B-r}), (\ref{LL-ISCO-BB}), and
(\ref{BB-ISCO-JJ}) at the horizon $r=2$ and assuming that the
neutral particle starts its motion at the rest at infinity with
the energy given by (\ref{enerflat}) as $\EE_n = \sqrt{1+\JJ^2}$,
one can obtain the collisional energy per unit mass $(\EE_{\rm
c.m.} = E_{\rm c.m.}/m)$ in the following form:
\begin{equation}
\EE_{\rm c.m.}^{\rm max} \approx 1.74 \left(\BB
\sqrt{1+\JJ^2}\right)^{1/4}. \label{cmfinal}
\end{equation}

\section{Rotating black string as particle accelerator\label{secIV}}

\subsection{Dynamical equations}

As it is generally known, the metric of the Kerr black hole
spacetime is the four-dimensional rotating neutral axially
symmetric solution of the vacuum Einstein equations. When the
additional dimension $w$ is included to the metric of the Kerr
spacetime in the Boyer-Lindquist coordinates then the obtained
solution takes the following form \cite{Horowitz2002,Grunau:2013}
\begin{eqnarray}
 \dd s^2& =& \frac{\rho^2}{\Dr}\dd r^2 + \rho^2\dd\vartheta^2 + \dd
 w^2 -  \frac{\Dr}{\rho^2}(\dd t-a\sin^2\vartheta\dd\varphi)^2 +{}
 \nonumber\\
 &&\frac{\sin^2\vartheta}{\rho^2}\left[(r^2+a^2)\dd\varphi-a\dd
 t\right]^2  ,\label{spacetimerbs}
\end{eqnarray}
where the metric parameters $\Dr$
and $\rho$ are defined as
\begin{equation}
\Dr= r^2-2Mr+a^2, \qquad \rho^2=r^2+a^2\cos^2\vartheta.
\end{equation}

Eq.~(\ref{spacetimerbs}) is the metric of the spacetime of a
uniform rotating black string. Here $M$ is the mass density of the
black string and $a$ is the parameter of the rotation of the black
string with the relation $0\leq a/M\leq1$. The singularity, as
usually, placed at $\rho^2=0$, (or at $r=0$ and
$\vartheta={\pi}/{2}$). This implyies, that as in the Kerr
{geometry}, the motion of a particle along the geodesic line with
$r=0$ and $\vartheta\neq{\pi}/{2}$ will not end at the point of
singularity, {and} the negative values of $r$ are allowed in the
rotating black string spacetime. \cite{ONeill:1995,Grunau:2013}.
The expression $\Dr=0$, allow us to define the event horizons of
the black string as
\begin{equation}
 r_\pm = M \pm \sqrt{ M^2 - a^2}  ,
\end{equation}
which coincide with the event horizon in the Kerr spacetime.

 Associated with a
timelike $\xi^{\mu}_{(t)}$ and spacelike $\xi^{\mu}_{(\varphi)}$
and $\xi^{\mu}_{(w)}$ Killing vectors one can find the following
conserved quantities along a geodesic {line} on the equatorial
plane $\vartheta=\pi/2$ as:
\begin{eqnarray}
\EE &= &\frac{E}{m} = -\xi^{\mu}_{(t)} u_{\mu} =
\left(1-\frac{2M}{r}\right) \frac{dt}{d\tau} + \frac{2 M a}{r}
\frac{d\varphi}{d\tau}, \label{intE}
\\
\LL &=&\frac{L}{m} = \xi^{\mu}_{(\varphi)} u_{\mu} =
\left(r^2+a^2+\frac{2 M a^2}{r}\right)\frac{d\varphi}{d\tau} -
\frac{2 M a}{r} \frac{dt}{d\tau}, \label{intL}
\\
\JJ &=& \frac{J}{m} = \xi^{\mu}_{(w)} u_{w} = \frac{d w}{d\tau}.
\label{intJb}
\end{eqnarray}

\begin{figure}[h]
 \centering
   \includegraphics[width=7cm]{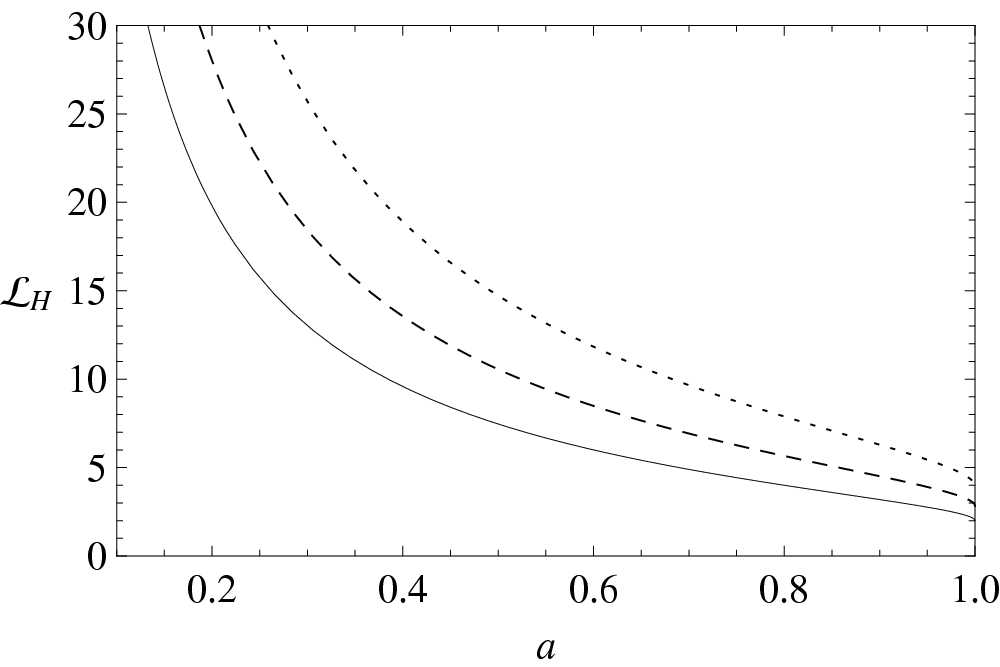}
 \caption{The limiting value of the angular momentum of the
particle near the horizon $r=r_+$ of the rotating black string
with dependence on the spin parameter $a$ for the different values
of the parameter $\JJ$: $J = 0$ (solid), $J = 1$ (dashed), $J =
1.7$ (dotted). The area of the possible values of the angular
momentum is shaded.} \label{Lrbs_Lh_a}
\end{figure}

\begin{figure*}
\subfigure[ \quad TO
$\LL=2$]{\includegraphics[width=0.3\hsize]{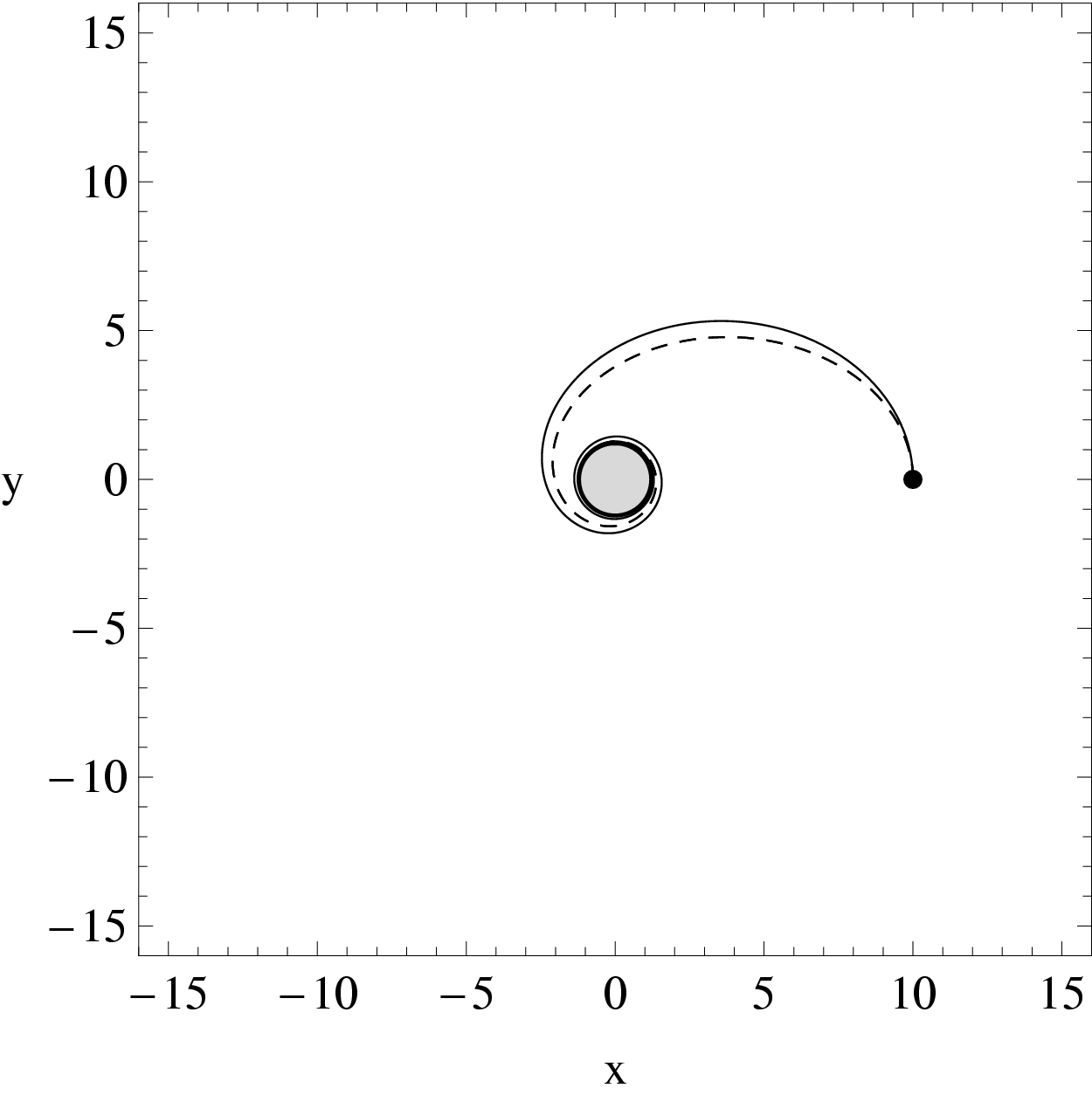}} \subfigure[
\quad BO $\LL=3$]{\includegraphics[width=0.3\hsize]{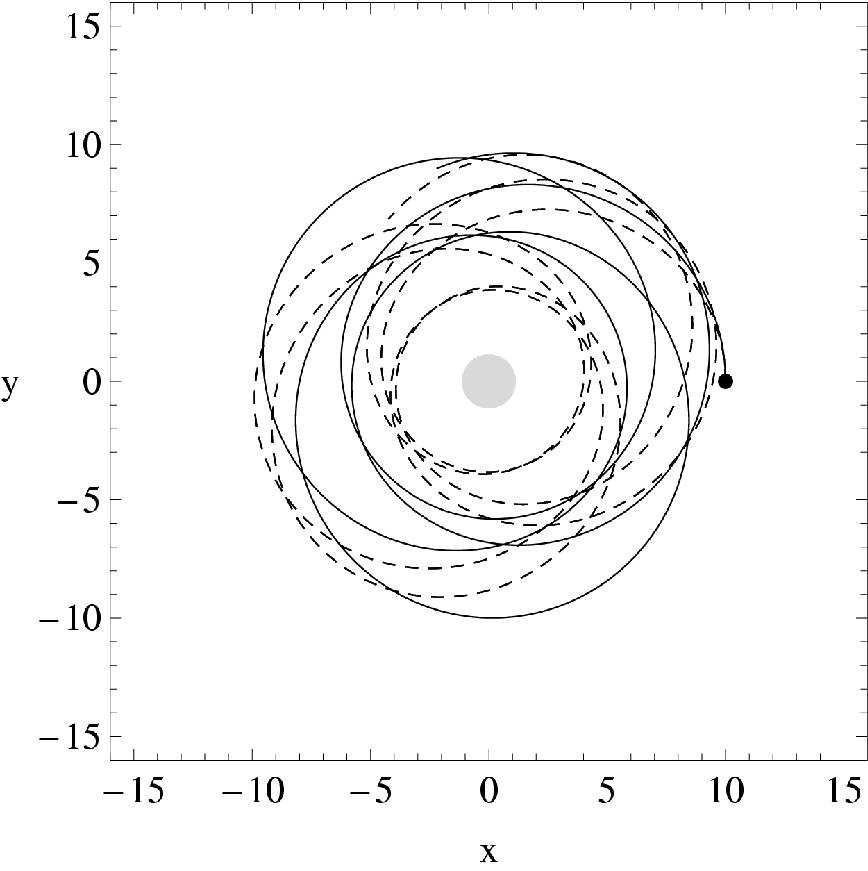}}
\subfigure[ \quad EO
$\LL=6$]{\includegraphics[width=0.3\hsize]{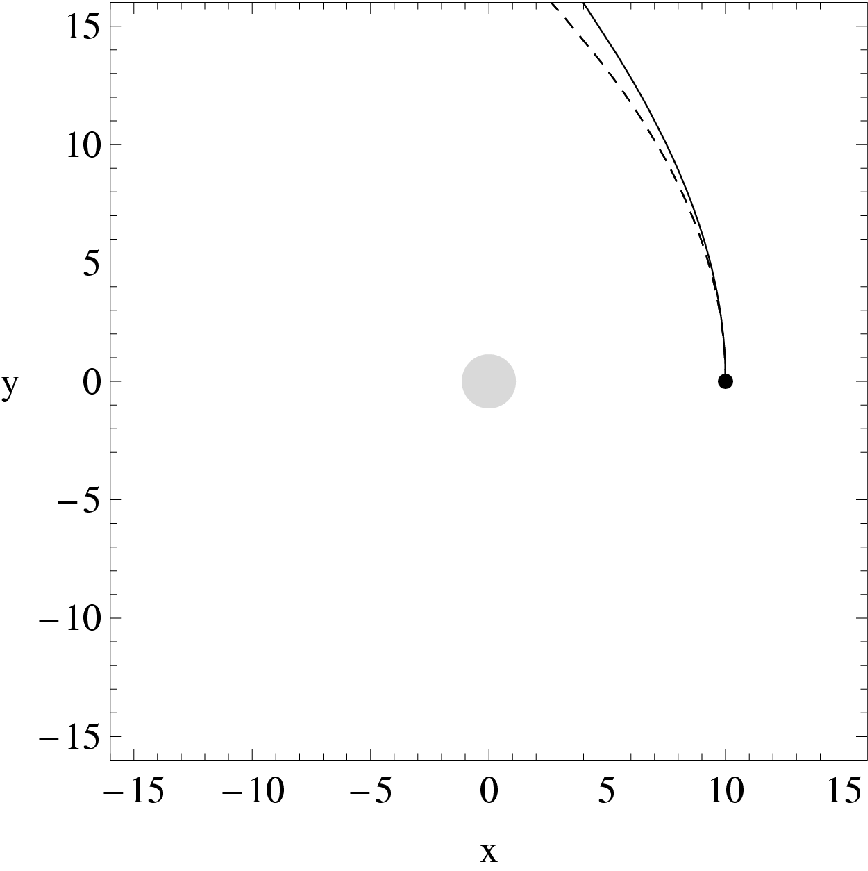}}
\caption{Examples of particle trajectories moving in equatorial
plane around rotating black string when the spin parameter
$a=0.99$. We compare motion with ${\cal J}=0$ (solid curve) and
with ${\cal J}=0.5$ (dashed) for all TO, BO and EO possible
orbits. Particles start from the initial position $r_0=10$ but
with the different energies.   \label{figNO2}}
\end{figure*}

\begin{figure*}
\includegraphics[width=\hsize]{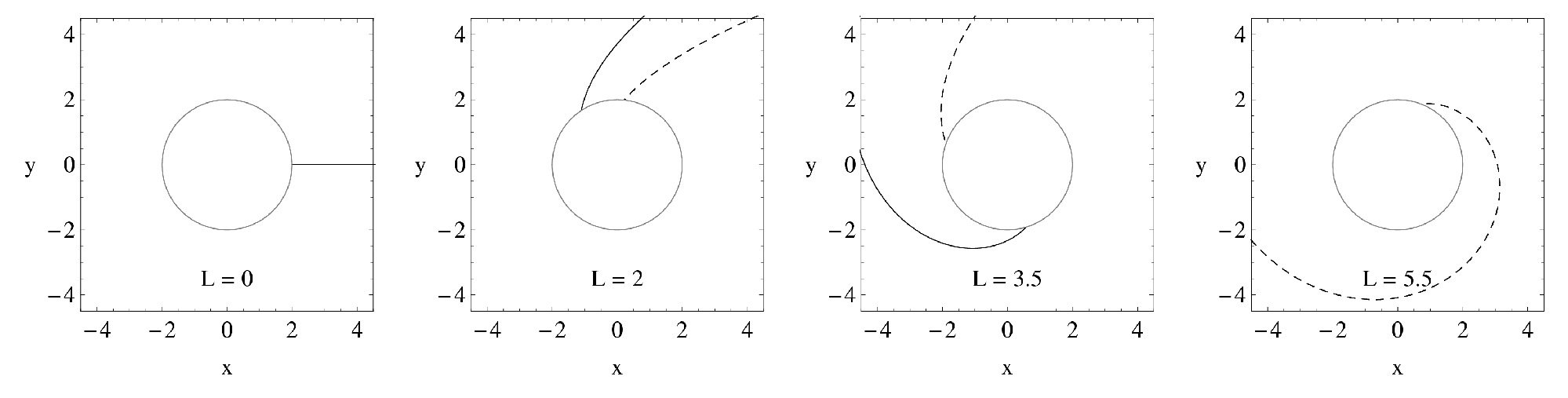}
\includegraphics[width=\hsize]{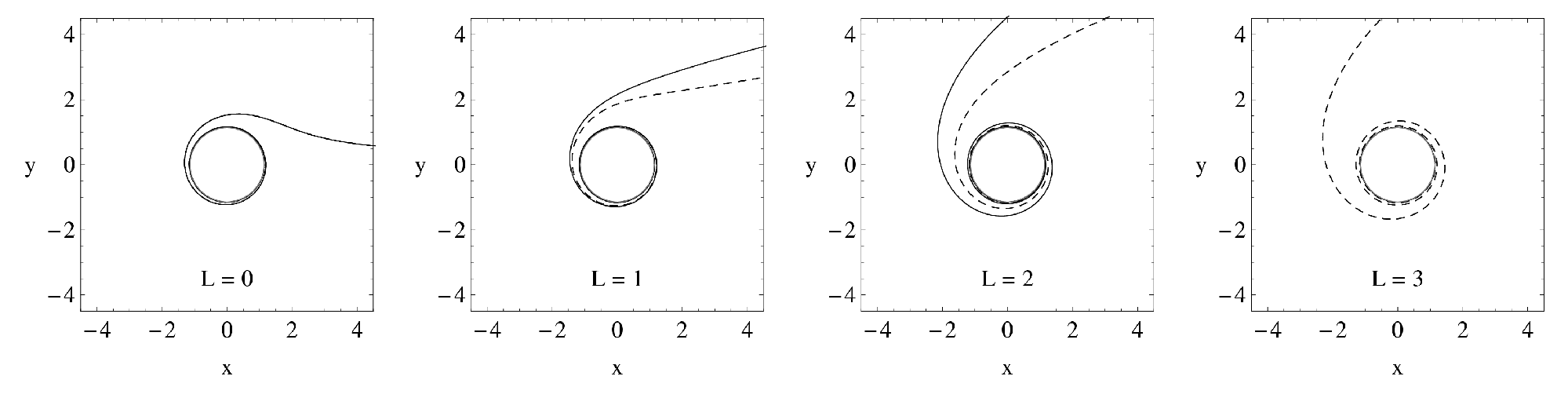}
\caption{We compare TO for particle motion with $\JJ=0$ (solid
curve) and with $\JJ=1$ (dashed one), depicting dependence of
"ending angle" $\varphi_{\rm end}$ from the specific angular
momenta $\LL$ for the case when the spin parameter is $a=0$ and
$a=0.99$. Particles start from the "infinity" (at $r_0=1000$) with
energies $E\doteq~1$ (for $\JJ=0$) and $\EE\doteq~1.4$ (for
$\JJ=1$). For some critical values of angular momentum $\LL$ the
TO orbits become BO and hence are not visible on the graphs. Such
critical value $\LL_{\rm c}$ is growing with parameter $\JJ$ (For
the case when $a=0$ we have $\LL_{\rm c} = 4\sqrt{1+\JJ^2}$.). We
also see strong "winding" of particle trajectories onto event
horizon of black string when  $a=0.99$. Even orbits with $\LL=0$
have nonzero "ending angle" $\varphi_{\rm end}$. \label{figNO3}}
\end{figure*}

The Hamilton-Jacobi equation \eqref{eqn:ham-jac} can be separated
and gives a differential equation for each component as
\cite{Grunau:2013}
\begin{eqnarray}
\frac{dt}{d\tau} &=& \frac{1}{\Delta
r}\left[\EE(r^3+2Ma^2+a^2r)-2aM\LL\right], \label{rot-t-eq}
\\
\frac{d\varphi}{d\tau}& =& \frac{1}{r
\Delta}\left[2aM\EE+(r-2)\LL\right], \label{rot-phi-eq}
\\
\frac{d w}{d \tau} &= &\JJ, \label{rot-w-eq}
\\
\left(\frac{d r}{d \tau}\right)^2 &= &\EE^2 - V_{\rm eff},
\label{rot-r-eq}
\end{eqnarray}
where the effective potential has a form
\begin{eqnarray}
V_{\rm eff} &=&\frac{1}{r^2} [\LL^2\left(1-\frac{2M}{r}\right) - a^2
\EE^2\left(1+\frac{2M}{r}\right) +{}
\nonumber\\
 && 2a\EE \LL \frac{2M}{r}
+\Delta\left(1+\JJ^2\right)]. \label{veff}
\end{eqnarray}
For getting the solutions of the equations of motion outside the
event horizon one has to hold the condition $dt/d\tau
>0$. At the region close to the horizon $r \rightarrow r_{\rm +}$, this condition
gives the limiting value of the angular momentum of the
particle as
\begin{equation}
\LL \leq \LL_{H} = \frac{2\EE}{a} \left(1+\sqrt{1-a^2}\right).
\label{lhlimit}
\end{equation}
Equation (\ref{lhlimit}) is similar for both Kerr black hole and
rotating black string {spacetimes}. However, for the particle
falling on the horizon with the initial rest energy
(\ref{enerflat}), the limiting (maximal) value of the angular
momentum takes the form
\begin{equation}
\LL_{H} = \frac{2}{a} \left(1+\sqrt{1-a^2}\right) \sqrt{1+\JJ^2} .
\label{lhlimitJ}
\end{equation}
The variation of the function $\LL_H$ with dependence on the spin
parameter $a$ for the different values of the parameter $\JJ$ is
shown in Fig.\ref{Lrbs_Lh_a}. It {is} easy to see that when the
value of the parameter $\JJ$ increases, the limiting allowed value
of the angular momentum also increases. Samples of particle
trajectories moving in equatorial plane around rotating black
string are shown in Fig.~\ref{figNO2} for the different values of
parameter ${\cal J}$ when the spin parameter $a=0.99$, depicting
the change from TO to BO and EO orbits. Such transitions is shown
in Fig.\ref{figNO3} for $a = 0$ and $a = 0.99$.

\subsection{Freely falling particle}

In this subsection, we study the properties of the freely falling
{particles} into the rotating black string and approaching the
horizon with the different values of momenta ${\cal L}_1$, ${\cal
J}_1$ and ${\cal L}_2$, ${\cal J}_2$ at the equatorial plane
$\vartheta=\pi/2$ of the black string.

\begin{figure*}[t!]
 \includegraphics[width=7cm]{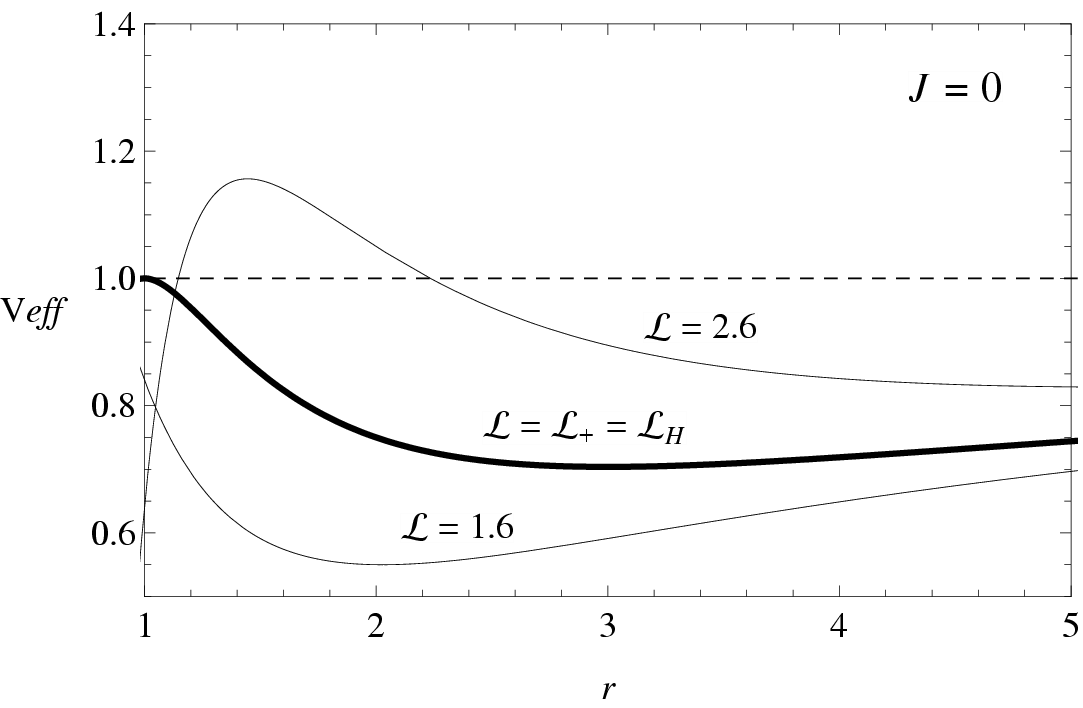}  \label{pic:effective-ext-rbs-a}
\includegraphics[width=7cm]{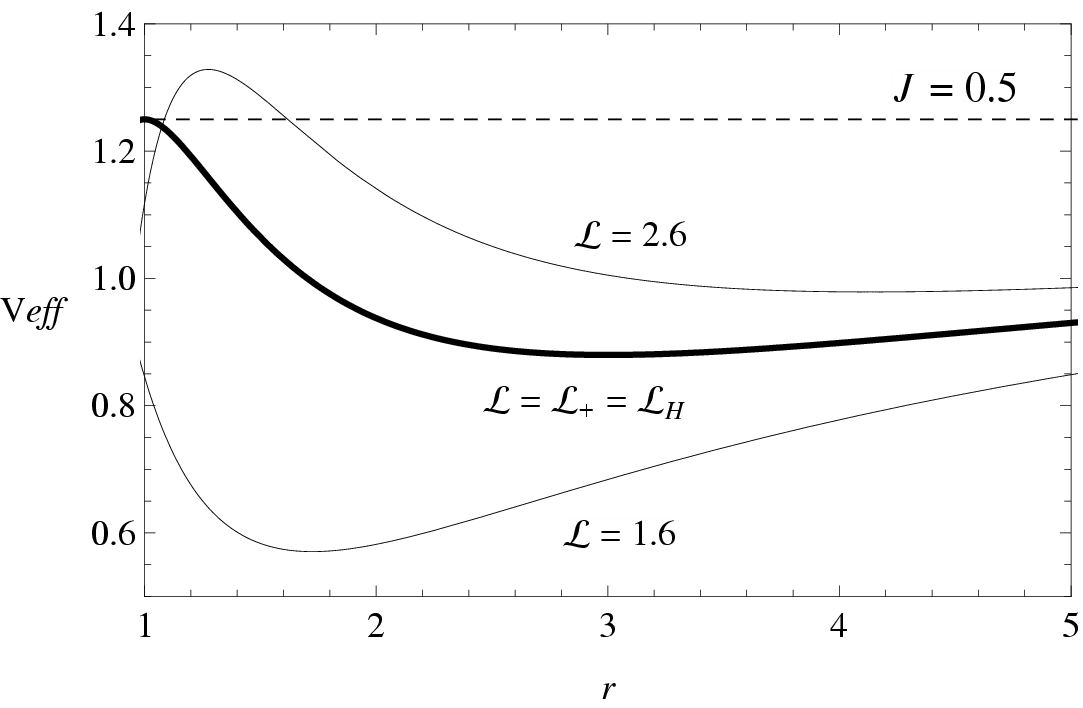}  \label{pic:effective-ext-rbs-b}
 \caption{The radial dependence of the effective potential of the particle
 moving around extremely rotating black string with $a=1$ for the different values of the angular
 momentum $\LL$ in two different cases of the absence (left panel) and presence (right panel) of the parameter $J$.
 Dashed lines indicate the energy level of a free particle at infinity.}
 \label{pic:effective-extrem-rbs}
\end{figure*}

\begin{figure*}[t!]
    \includegraphics[width=5cm]{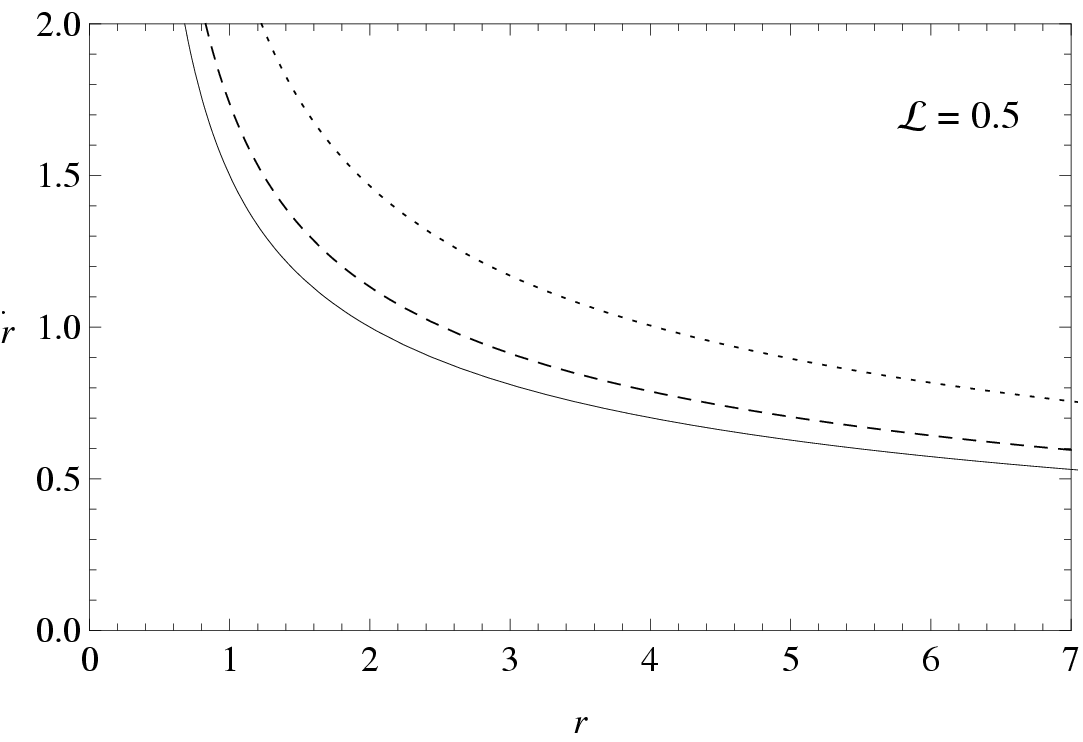}  \label{pic:acceleration-rbs-a}
    \includegraphics[width=5cm]{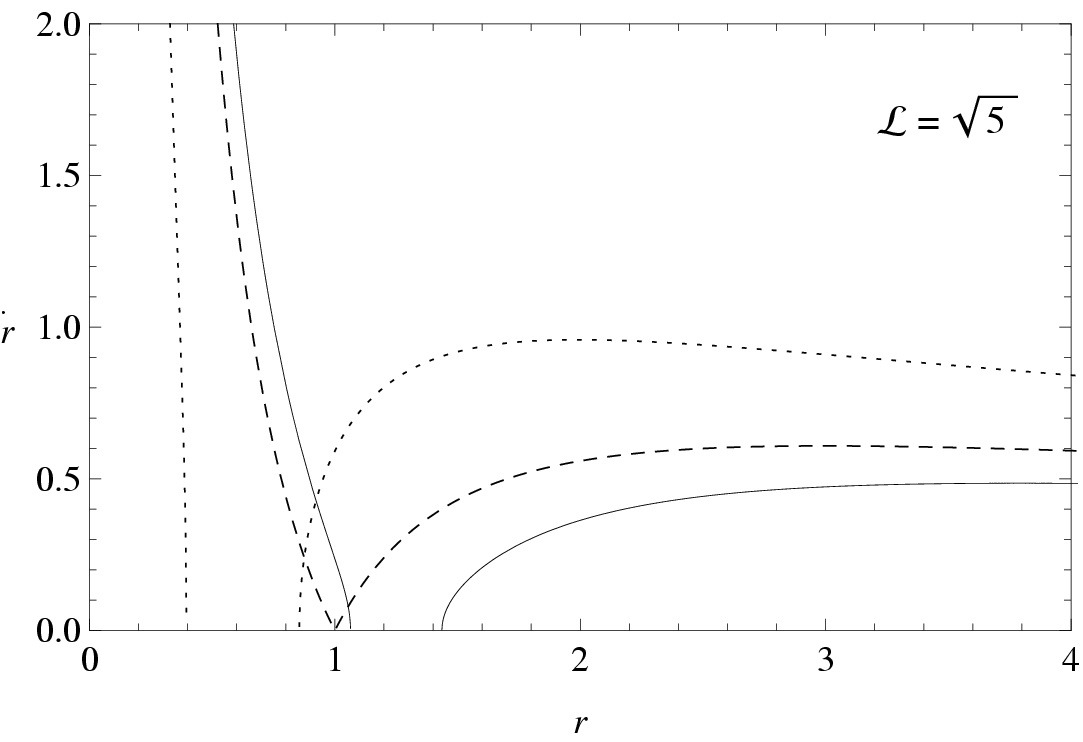}  \label{pic:acceleration-rbs-b}
    \includegraphics[width=5cm]{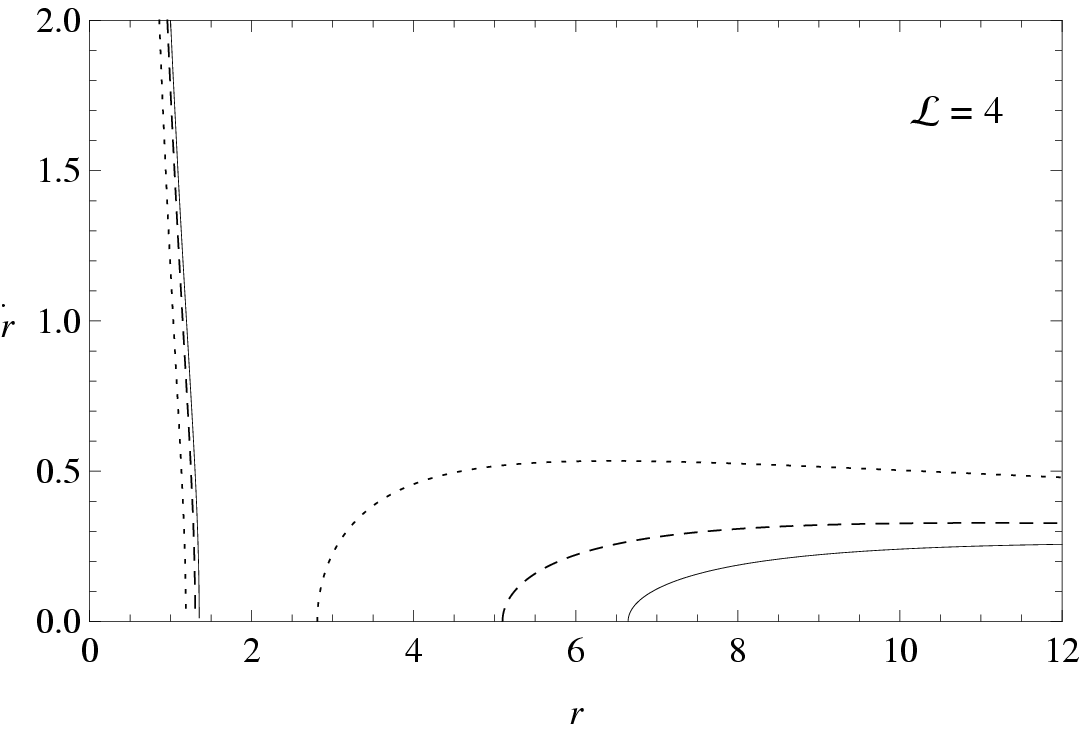}  \label{pic:acceleration-rbs-c}
 \caption{Acceleration of particle along $r$ by extremely rotating black string
  for three different values of the parameter $\JJ$: $\JJ=0$ (solid), $\JJ=0.5$ (dashed) and $\JJ=1$ (dotted) and three different values
 of the angular momentum: $\LL=0.5$ (left panel),  $\LL \approx 2.24$ (middle panel) and $\LL=4$ (right panel).
 Critical case corresponds to the dashed curve in the middle panel.}
 \label{pic:acceleration-rbs}
\end{figure*}

It is very important to find the limitations on the possible
values of the integrals of motion $\LL$ and $\JJ$ of particles to
achieve the horizon of the rotating black string as we did in
Sec.\ref{freeparSBS}. In the center-of-mass frame the maximal
energy will {take} place when the particles collide near a
rotating black string with maximal and opposite {momenta} $\LL_1$,
$\LL_2$ and {parameters} $\JJ_1$, $\JJ_2$. But if we set the large
values {for} them, then the geodesics may never reach the vicinity
of the horizon (see the plots with $\LL=1.6$ in
Fig.~\ref{pic:effective-extrem-rbs}) and effects of the strong
gravitational field cannot be analyzed. On the other hand, for the
small values of angular momentum, the particles fall radially with
a small tangential velocity and the center-of-mass energy does not
grow either (see the plots with $\LL=2.6$ in
Fig.~\ref{pic:effective-extrem-rbs}). In spacetime of rotating
black string, the particles approaching from one side or the other
have different properties which are similar to their properties in
Kerr spacetime. Consequently, there are critical values for the
parameters $\LL$ and $\JJ$ such that particles may reach the
horizon with maximum tangential velocity (see thick curves in
Fig.~\ref{pic:effective-extrem-rbs}). To find these values we use
Eq.~(\ref{rot-r-eq}) and its derivative. The profiles of the
variation of $\dot{r}$ with radius for the different values of the
parameters $\JJ$ and $\LL$ are shown in
Fig.\ref{pic:acceleration-rbs}, where the critical case when the
particle approaches the horizon of the black string is indicated
by the dashed curve of the plot ($b$).

In order for the massive particle freely falling with momentum
$\JJ$ to the black string with the spin parameter $a$ to achieve
the horizon, it must have the angular momentum in the range
\begin{equation}
\LL_{-} \leq \LL \leq \LL_{+}, \label{L-crit-rbs}
\end{equation}
where $\LL_{+}$ and $\LL_{-}$ {are defined as}
\begin{equation}
\LL_{+} = 2 \sqrt{1 + \JJ^2}\left(1 + \sqrt{1 - a}\right),
\label{L-crit-rbs-plus}
\end{equation}
\begin{equation}
\LL_{-} = - 2 \sqrt{1 + \JJ^2}\left(1 + \sqrt{1 + a}\right).
\label{L-crit-rbs-minus}
\end{equation}
%

\begin{figure*}[t!]
 \includegraphics[width=7cm]{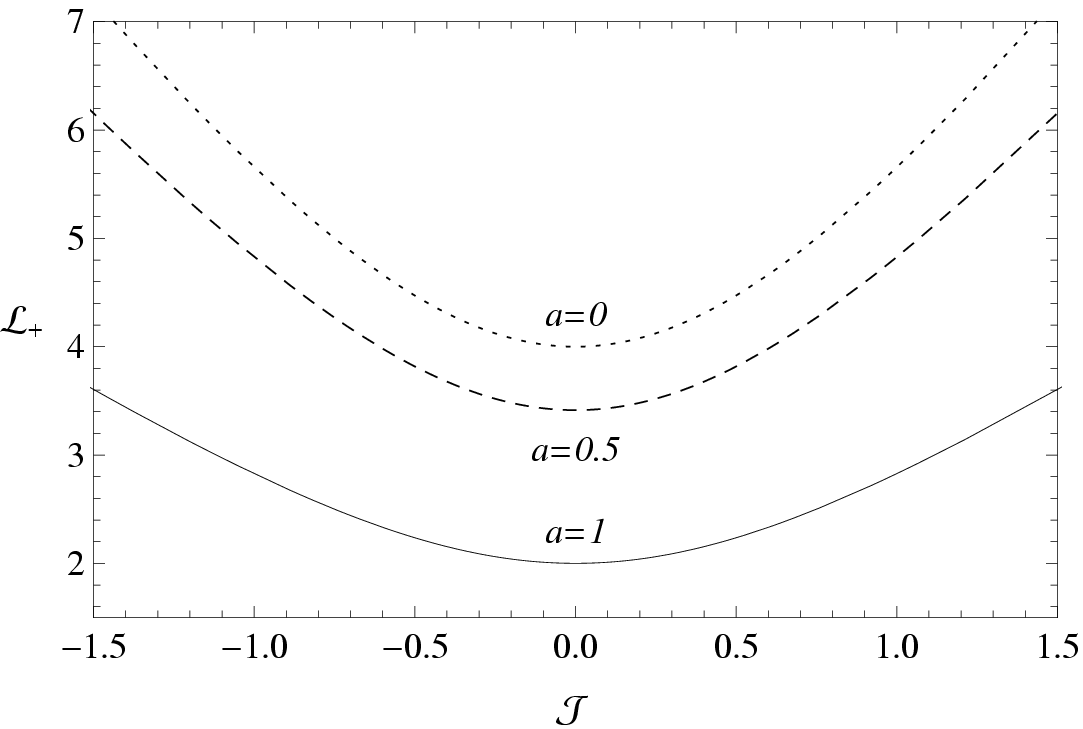} \label{L_crit_J_plus}
 \includegraphics[width=7cm]{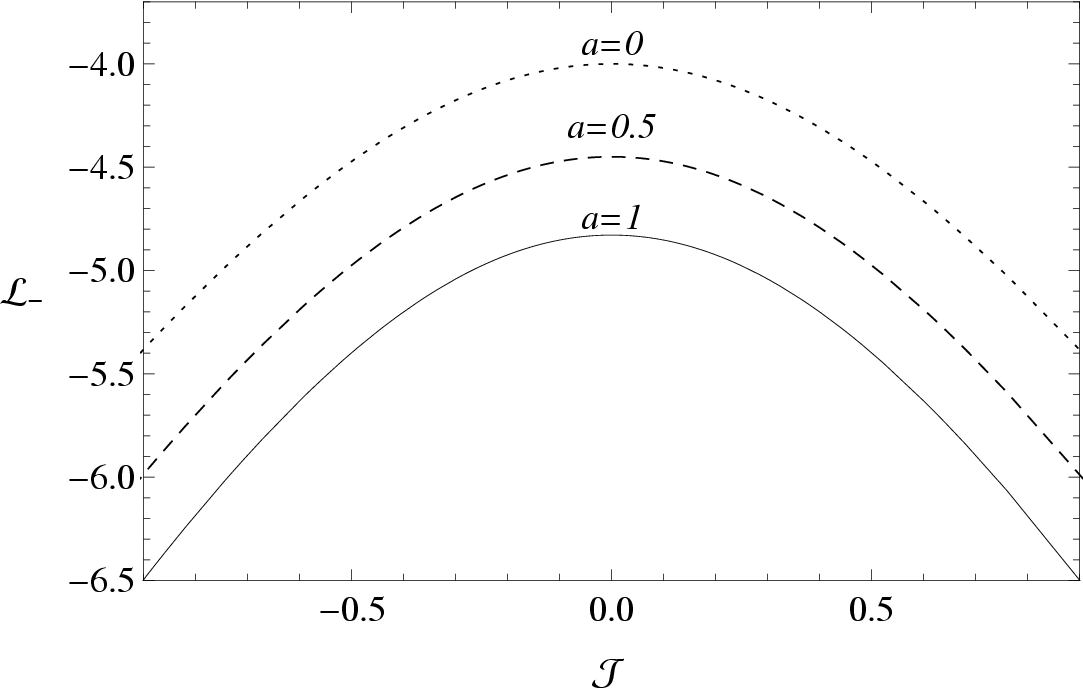} \label{L_crit_J_minus}
 \caption{Dependence of the critical angular momentum $\LL_{\pm}$ of the particle on the parameter $\JJ$
 for the different values of the spin parameter $a$ of the black string. }
 \label{L_crit_rbs}
\end{figure*}

It is important to note that this does not apply when the particle
scatters with other particles and changes its energy and angular
momentum on the way to the horizon. The dependence of the critical
angular momenta $\LL_{\pm}$ of the particle reaching the horizon
of the extremal black string on the values of {parameter} $\JJ$
are shown in Fig.\ref{L_crit_rbs}. It can be seen that the
solutions of the equation for the particle moving in the spacetime
of black string are invariant under the reversal of sign of
{parameter} $\JJ$. When the extra dimension is added, the absolute
values of the critical angular momenta increase for any value of
the spin parameter $a$.

\subsection{Collision of freely falling particles near extremely rotating black
string}

In this subsection we study the collision of two particles
approaching the horizon of the extremely rotating black string
with the different momenta ${\cal L}_1$, ${\cal J}_1$ and ${\cal
L}_2$, ${\cal J}_2$ at the equatorial plane $\vartheta=\pi/2$ of
the black string.

First, we derive the general formula for the center-of mass energy
of the collision of two particles freely falling from the spatial
infinity to the horizon of the rotating black string. Considering
the center-of-mass frame and a pair of particles with the same
mass $m_0$ and velocities represented by $U_{(m)} = U^\mu_{(m)}$,
one can write the expression for the energy of the collision of
two particles in the form
\begin{eqnarray}
E_{\rm c.m.}^2 = \left(m U^{\alpha}_{(1)} + m
U^{\alpha}_{(2)}\right) \left(m U_{(1)\alpha} + m
U_{(2)\alpha}\right) {}
\nonumber\\
 = 2 m^2 \left(1+U^{\alpha}_{(1)} U_{(2) \alpha}\right),
\end{eqnarray}
or in the form similar to (\ref{encolstat}) as
\begin{equation}
E_{\rm c.m.} = m_0 \sqrt{2}\sqrt{1-g_{\alpha\beta} U^\alpha_{(1)}
U^\beta_{(2)}},  \label{engenfor}
\end{equation}
where $U^\alpha_{(1)}$ and $U^\beta_{(2)}$ are the velocities of
the particles, properly normalized by $g_{\alpha\beta} U^\alpha
U^\beta = -1$. Direct substitution of Eqs. (\ref{rot-t-eq}) -
(\ref{rot-r-eq}) to (\ref{engenfor}) gives the expression for the
energy of the collision of two particles $(\EE_{\rm c.m.} = E_{\rm
c.m.}/m_0)$
\begin{eqnarray}
\EE_{\rm c.m.}^2 &=& \frac{2}{r\Delta} \bigg\{ a^2 [\EE_1 \EE_2 (2
+ r) + r (1 - \JJ_1 \JJ_2)]  - {}
\nonumber\\
 && 2 a (\EE_2 \LL_1 + \EE_1 \LL_2) +
 r^2 [(1 - \JJ_1 \JJ_2) (r - 2) + \EE_1 \EE_2 r]
\nonumber\\
 &&- {}\LL_1 \LL_2 (-2 + r) - \sqrt{R_1}\sqrt{R_2}\bigg\}, \label{cmrbs}
\end{eqnarray}
colliding at some radius $r$ at the equatorial plane which is the
generalization of Eq.(\ref{ecmschw}) to the case of the rotating
black string spacetime.

{Here}
\begin{equation}
R_i = 2 (a \EE_i - \LL_i)^2 - \LL_i^2 r + 2 r^2 \LL_i^2, \qquad
i=1, 2\ ,
\end{equation}
and
\begin{equation}
\EE_i = \sqrt{1+\JJ_i^2} \qquad i=1,2
\end{equation}
for the freely falling particles.

\begin{figure*}[t!]
 \includegraphics[width=7cm]{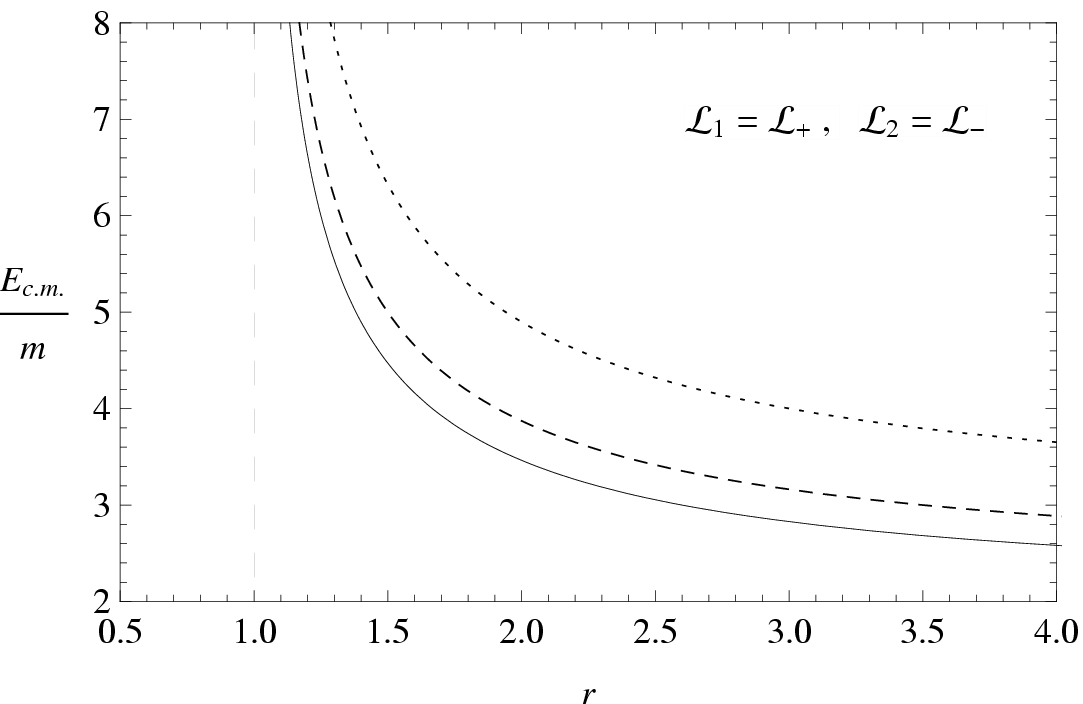}
   \includegraphics[width=7cm]{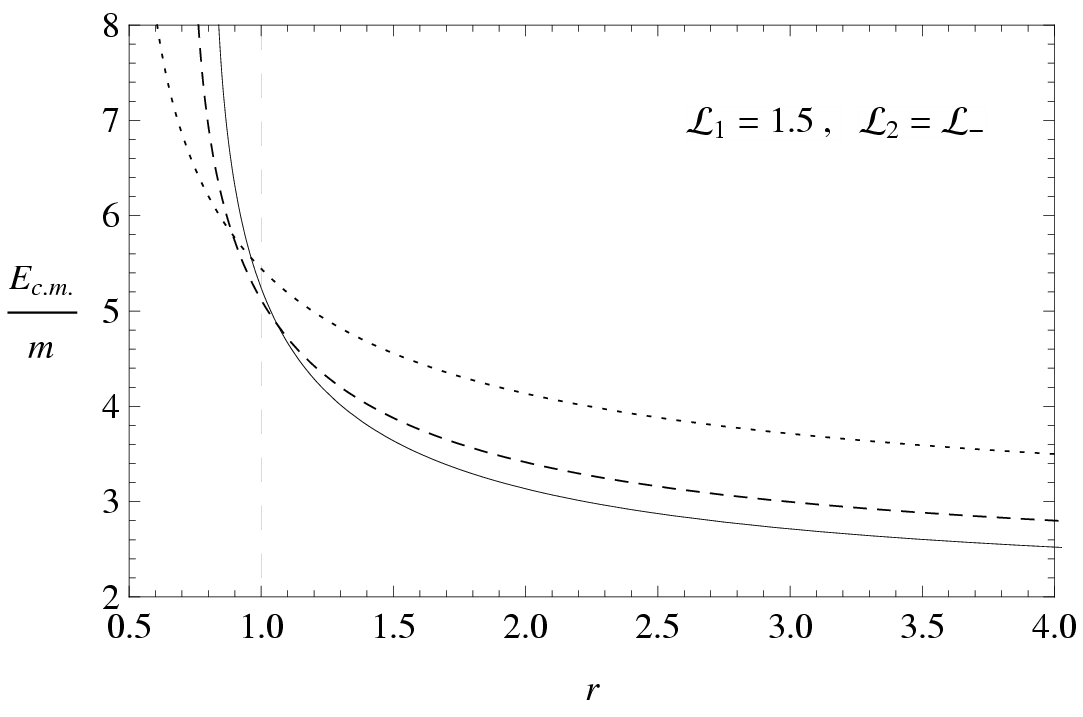}
 \caption{The variation of the energy of the collision of two particles with opposite
 momenta near the static black string in the center-of-mass frame for three different
 values of parameter $\JJ$. $\JJ_{1,2}=0$ (solid), $\JJ_{1,2}=\pm 1$ (dashed),
  $\JJ_{1,2}=\pm 1.5$ (dotted). The event horizon is indicated by vertical thin line. }
 \label{pic:ecm_rbs_r}
\end{figure*}

Ex facte, one can say that Eq.~(\ref{cmrbs}) diverges at the
horizon of the black string given by $\Delta=0$ and extremal
collisional energy appears. However, it is not satisfied (see
Fig.~\ref{pic:ecm_rbs_r}) since the numerator of Eq.~(\ref{cmrbs})
also vanishes at this point. This result corresponds to the result
of the paper \cite{BSW:2009} in the case when $\JJ_i = 0$.

To find a limit of the energy at the horizon $r\rightarrow r_{+}$
one may use the l'Hospital's rule and find the actual extremal
collisional energy at the horizon of extremely rotating black
string with $a=1$. Thus, the energy of the collision in the
special case with $\JJ_1=-\JJ_2=\JJ$ (which is relevant due to the
symmetry of the additional dimension) will have the following form
\begin{eqnarray}
\EE_{\rm c.m.}^2 |_{(r\rightarrow r_+)} &=& 2(1+\JJ^2)
 \label{ecmrhaex}\\&\times&\left(\frac{\LL_1
- 2\sqrt{1 + \JJ^2}}{ \LL_2 - 2\sqrt{1 + \JJ^2}} + \frac{ \LL_2 -
2\sqrt{1 + \JJ^2}}{\LL_1 - 2\sqrt{1 + \JJ^2}}\right). \
\ \nonumber
\end{eqnarray}
Expression (\ref{ecmrhaex}) diverges when the dimensionless
angular momentum of one of the colliding particles $\LL=\LL_+$ and
on the contrary finite for the generic values of $\LL_1$ and
$\LL_2$. Thus for extremely rotating black strings the scattering
energy in the center-of-mass frame becomes unlimited when
$\LL=\LL_+$.

This shows that a singularity in the center-of-mass energy is
achieved on the extremal horizon for most specific values of
angular momentum and is finite for the generic values of
{momentum} $\LL$. In this case, every value of the finite energy
is achieved up to the event horizon and infinite center-of-mass
energy is obtained only for particle collisions {occurring} on the
horizon as shown in Ref.~\cite{BSW:2009}. If the angular momentum
of both individual particles is greater than the critical one,
then they will never reach the horizon. Vise versa, if they both
have angular momentum below the critical value, then they will
fall into the black string with a finite center-of-mass
collisional energy.

\subsection{The collision energy near nonextremely rotating black string}

It has been recently shown in Ref.\cite{Sij-Chan:2011:PRD} that an
arbitrarily high energy of the collision of two freely falling
particles near a Kerr black hole cannot occur if a spin parameter
$a$ is not maximal. Similarly, in the spacetime of a rotating
black string as we will show below, there is no possibility to
obtain an unlimited energy from the collision of freely falling
from infinity particles if the rotation of the black string is not
extremal. {However}, in the processes {accompanied by} the
multiple scattering when the particles fall into the black string
and collide with other particles moving around a horizon of the
black string high energy collision can be obtained even if the
rotation of the black string is not maximal, i.e., $a<1$. In this
subsection we study the possibility of the extraction of the
unlimited energy from the particles collision for the case of a
nonextremal black string.
\begin{figure}[h]
   \includegraphics[width=7cm]{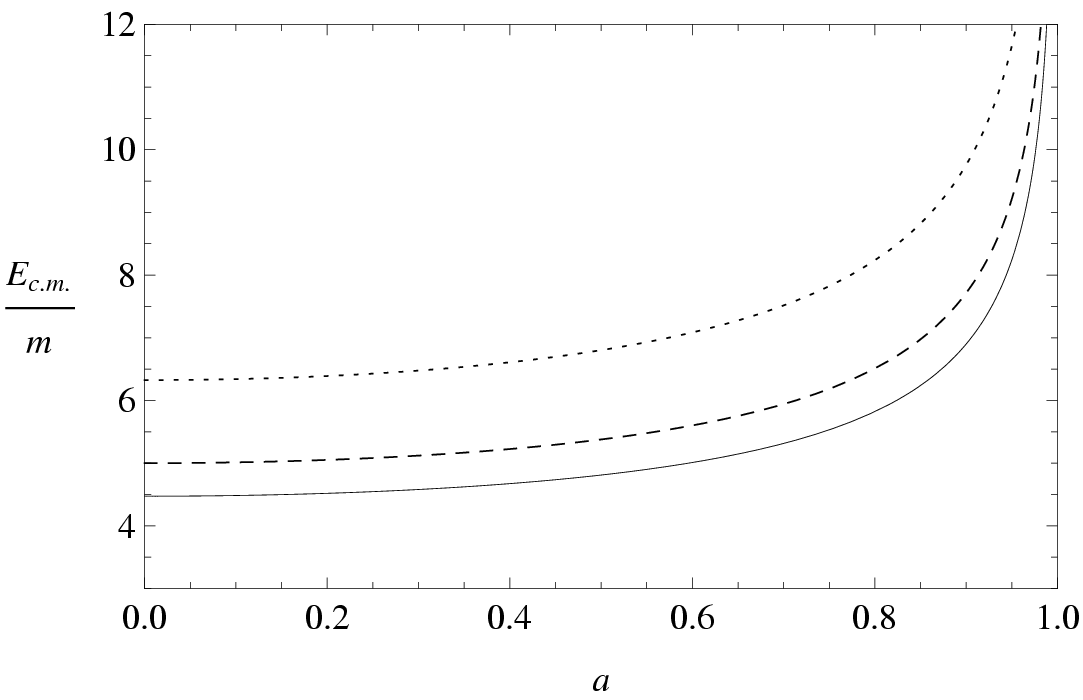}
 \caption{The center-of-mass energy of the collision of two freely falling particles
  close to the horizon of the rotating black string in dependence on the spin parameter $a$
  for three representative values of the parameter $\JJ_{1,2}$ of each particle: $\JJ_{1,2}=0$ (solid),
  $\JJ_{1,2}=\pm 0.5$ (dashed), and $\JJ_{1,2}=\pm 1$ (dotted). \label{Ecm_as_A} }
 \end{figure}

As mentioned above, Eq.~(\ref{rot-r-eq}) leads to an appearance of
the limitation on the possible values of the angular momenta of
freely falling particles to reach the horizon of the black string.
For the rotating black string, this condition is given by
Eq.(\ref{L-crit-rbs}). Using the limiting values of the angular
momenta (\ref{L-crit-rbs-plus}) and (\ref{L-crit-rbs-minus}) of
the particles at the horizon of the rotating black string one can
find the dependence of the center-of-mass energy on the spin
parameter $a$ which is shown in Fig.\ref{Ecm_as_A}. It can be seen
from Fig.\ref{Ecm_as_A} that the unlimited energy occurs when the
black string rotates with the maximal spin parameter $a=1$,
 while for a black string with spin parameter
$a<1$, there will be an upper bound to the energy. It is very
interesting to find how the largest energy of the collision grows
as the maximally spinning case is approached. The maximal energy
{can be estimated} with the help of Eq.(\ref{cmrbs}). In terms of
the small parameter $\epsilon = 1 - a$, the main contribution to
the energy of the collision per unit mass is
\begin{equation}
\EE_{\rm c.m.}^{\rm max} \approx \frac{4.06}{\sqrt[4]{\epsilon}}
\left[\left(1+\JJ_1^2)(1+\JJ_2^2\right)\right]^{1/4}.
\end{equation}
In particular, according to \cite{Thorne:1974} the maximal
possible value of the spin parameter $a$ of the astrophysical
black holes is $a=0.998$. If one applies this limiting value of
the spin parameter for the estimation of the collision energy of
two particles with $\JJ_1=\JJ=-\JJ_2$ near the horizon of the
rotating black string, then the maximal possible center-of-mass
energy takes the value
\begin{equation}
\EE_{\rm c.m.}^{ \rm max} = 18.971 \sqrt{1+\JJ^2}, \qquad a=0.998\
.
\label{ecmnonextremal}
\end{equation}
{It can be seen that} even for the values of the spin parameter
$a$ {being} close to the extremal $a = 1$ of the rotating black
string {energy}, $E_{\rm c.m.}/m$ cannot be unlimited. However, as
it has been already noted in Sec.\ref{secII} the expression
(\ref{ecmnonextremal}) for the collisional energy tends to
infinity if the parameter $\JJ$ also grows in this limit ($\JJ
\rightarrow \infty$) {and} formally one can obtain an arbitrarily
large energy. However this case does not correspond to our
purposes since this high energy is not the result of the collision
of particles near the horizon of the rotating black string but the
result of an {enormous increase of the} initial energies of the
particles measured at infinity.

\begin{figure*}[t!]
    \includegraphics[width=7cm]{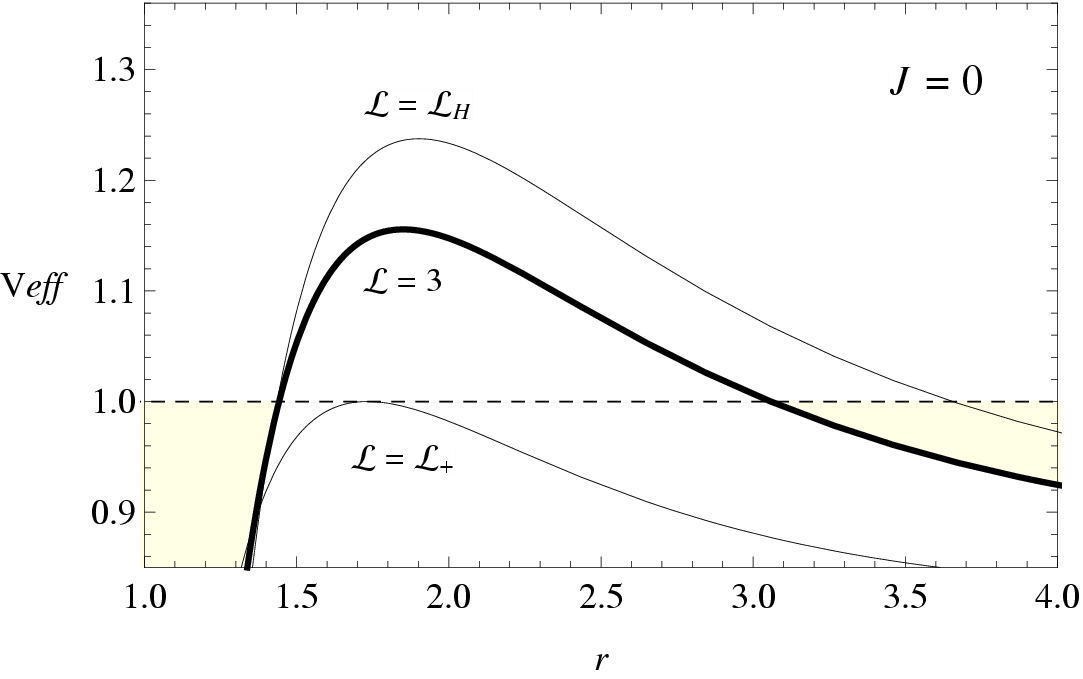}  \label{pic:effective-none-rbs-a}
    \includegraphics[width=7cm]{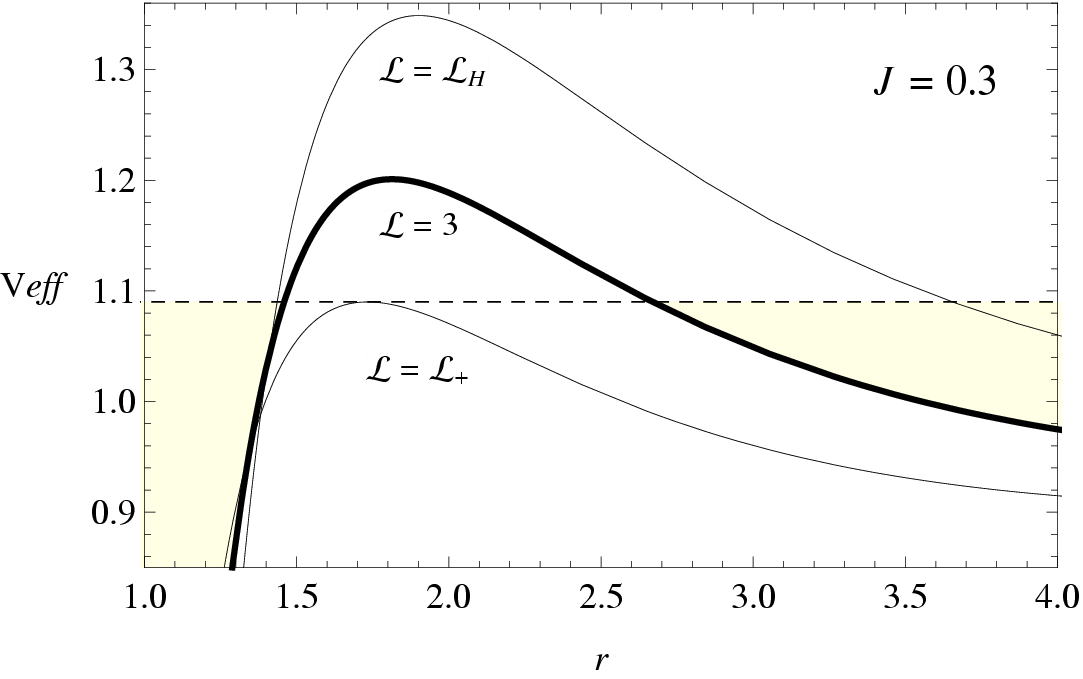}  \label{pic:effective-none-rbs-b}
 \caption{The effective potential of the particle
 moving around nonextremal black string with $a=0.9$ in the absence (left panel)
 and the presence (right panel) of the parameter $\JJ$. The allowed zone for $\LL=2.5$ is shaded.
 Dashed lines represent the energy level of a free particle at infinity. \label{pic:effective-none-rbs}}
\end{figure*}

 On the other hand, one may take into
account the possibility of a multiple scattering of the particle
falling from infinity with the fixed angular momentum. As the
result of interaction with the particles from the accreting disc,
the particle falling into the black string  can change its
momentum and again scatter close to the horizon, and {finally} the
scattering energy can be unlimited
\cite{Grib-Pavlov:2011:Grav-Cos}.

A particle moving from infinity to the black string can achieve
the horizon if its angular momentum is lying in the range $\LL_-
\leq \LL \leq \LL_+$ (see the dependence of $\LL_-$ and $\LL_+$
from $\JJ$ in Fig.~\ref{L_crit_rbs}). The effective potential of
the particle in the spacetime of nonextremely rotating black
string for the different values of $\LL$ and $\JJ$ is shown in
Fig.~\ref{pic:effective-none-rbs}. The particle can move in the
vicinity of the black string if its energy $\EE < \sqrt{1+\JJ^2}$.
Zones of allowed particle motion in the horizon vicinity are
shaded in Fig.~\ref{pic:effective-none-rbs}. One can see from
Fig.~\ref{pic:effective-none-rbs} that the presence of the new
constant of motion $\JJ$ shifts and increases the allowed zone of
particle motion. {However}, if the particle is going not from the
infinity but from a distance $r=r_H + e$ then due to the form of
the effective potential it may have the angular momentum of the
value $\LL = \LL_H - \delta$ larger than $\LL_+$ and fall on the
horizon. On the other hand if the particle falling from infinity
with $\LL \leq \LL_+$ achieves the same region $r=r_H + e$ and
interacts with other particles of the accretion disc or decays
into a lighter particle which results in an increased angular
momentum $\LL_1 = \LL_H -\delta$, then the main contribution to
the scattering energy in the center-of-mass frame according to
Eq.~(\ref{cmrbs}) takes the form
\begin{equation}
\EE_{\rm c.m.} \approx \frac{ \sqrt{2}}{\sqrt{\delta}}
\left(\frac{\LL_H -
\sqrt{1+\JJ^2}\LL_2}{1-\sqrt{1-a^2}}\right)^{\frac{1}{2}}
\left(1+\JJ^2\right)^{\frac{1}{4}}\ , \label{ecm-nonextrem}
\end{equation}
{if expanded in series.}

For the rotation parameter $a=0.998$ and $\LL_2=\LL_{-}$, the
expression (\ref{ecm-nonextrem}) reduces to the form
\begin{equation}
\EE_{\rm c.m.}^{\rm max} = \frac{3.854}{\sqrt{\delta}}
(1+\JJ^2)^{3/4}.
\end{equation}

 The energy (\ref{ecm-nonextrem}) grows without limit when
$\delta \rightarrow 0$. Note that for rapidly rotating black
strings, the difference between $\LL_H$ and $\LL_+$, e.g., for
$a=0.998$ and $\JJ=0.7$, $\LL_H - \LL_+ \approx 0.05$. The
possibility of getting small additional angular momentum in
interaction close to the horizon seems {to be very} probable.

\section{Efficiency of Energy Extraction\label{secV}}

Consider the geodesics of the  particles in
the equatorial plane $\vartheta=\pi/2$. The radial equation of
motion in the spacetime metric~(\ref{statmetric}) is given by
\begin{eqnarray}
\frac{m^2}{2}\left(\frac{dr}{d\tau}\right)^{2}
+
V(r)=0,
\label{radmot}
\end{eqnarray}
where
$m$ is the mass, $E$ is the energy, $L$ is the momentum of the particle, and
%
\beq
V(r)=\frac{\Delta}{2r^2}(m^2+J^2)+\frac{L^{2}-a^{2}E^{2}}{2r^{2}}
-\frac{M(L-aE)^{2}}{r^{3}}-\frac{E^{2}}{2} 
\label{effpotah}
\eeq
%
is the effective potential of radial motion of the particle.

Consider the particle escape
to infinity
analysing the
effective potential~(\ref{effpotah}).
Using the condition
$V(r)=0$
one can obtain
\begin{eqnarray}
b=
b_{\pm}(r)=
\frac{-2aM \pm r
\sqrt{\Delta(1-\kappa(1-2M/r)) }}{r-2M},
\end{eqnarray}
for the the
impact parameter
$b=L/E$
for the massless particle,
where  new dimensionless parameter
$\kappa=J/E$ is introduced.
This indicates that a massless particle with impact parameter
$b=b_{\pm} (r)$
will have  a turning point at the position
$r$.
When the black string is extreme rotating, i.e. $a=M$,
then
$b_{+}(r)$
starts from
$2M$ independently of the value of $\kappa$
and will increase to
the infinity with increasing of
radial coordinate  $r$.
$b_{-}(r)$
starts from
$2M$ {which} is larger than
$b_{+}(r)$
and will increases to the infinity with the increase of radial coordinate
$r$.
As radial coordinate increases above
$2M$
to the infinity
$b_{-}(r)$
starts from the
negative infinity and
increases to some
maximum
then monotonically
decreases to the
negative infinity.

\begin{table*}
\caption{\label{table1} The values of local maximum of the impact parameter $b_{-}$ for the different values of parameter $\kappa$. }
\begin{ruledtabular}
\begin{tabular}{|c|c|c|c|c|c|c|c|c|c|c|c|}

 %
  $\kappa$ & 0 & 0.1 & 0.2 & 0.3 & 0.4 & 0.5 & 0.6 & 0.7 & 0.8 & 0.9 & 1.0 \\
  $b_{-{\rm max}}/M $& -7.0 & -6.985 & -6.939 & -6.862 & -6.751 & -6.603 & -6.412 & -6.168 & -5.857 & -5.444 & -4.828 \\
  $r/M $ & 4.0 & 4.005 & 4.02 & 4.048 & 4.089 & 4.147 & 4.231 & 4.354 & 4.545 & 4.887 & 5.829
%
\end{tabular}
\end{ruledtabular}

\end{table*}

If
 $b_{-{\rm max}}<b<2M$ then the particle will escape
  to infinity if it is
moving outward initially.
If $b>2M$ or $b<b_{-{\rm max}}$ then
 the particle eventually will escape
to infinity independently of the sign
of the initial velocity, for the case when it is outside the
outer turning point.
If
$b=2M$ or
$b=b_{-{\rm max}}$
then
the particle will escape to infinity,
for the case when it is initially
moving outside the turning point.

\begin{figure*}
%
  %
  \includegraphics[width=0.45\textwidth]{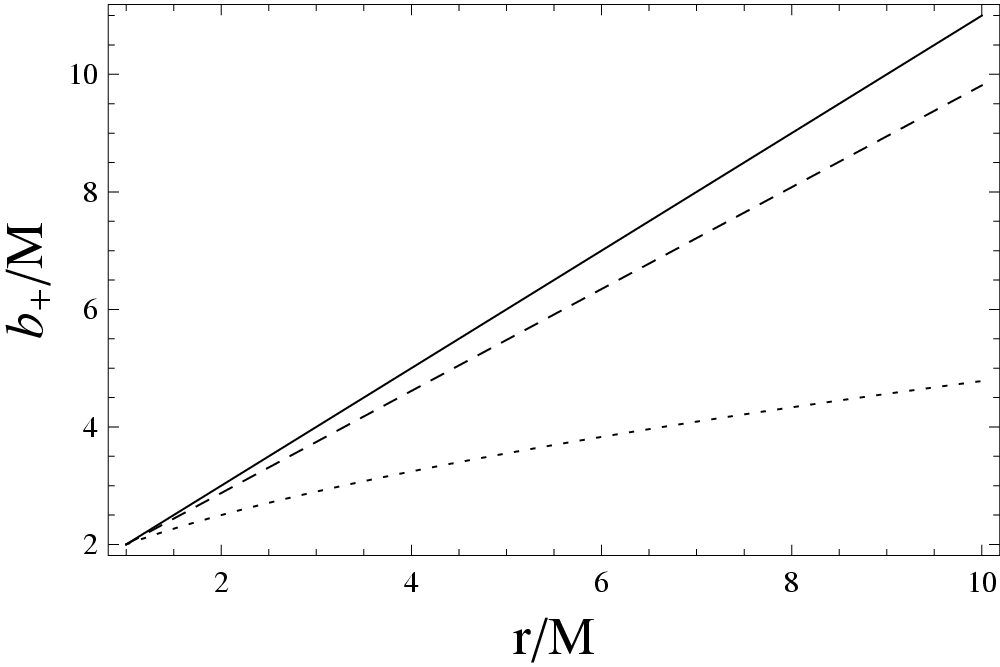} \includegraphics[width=0.45\textwidth]{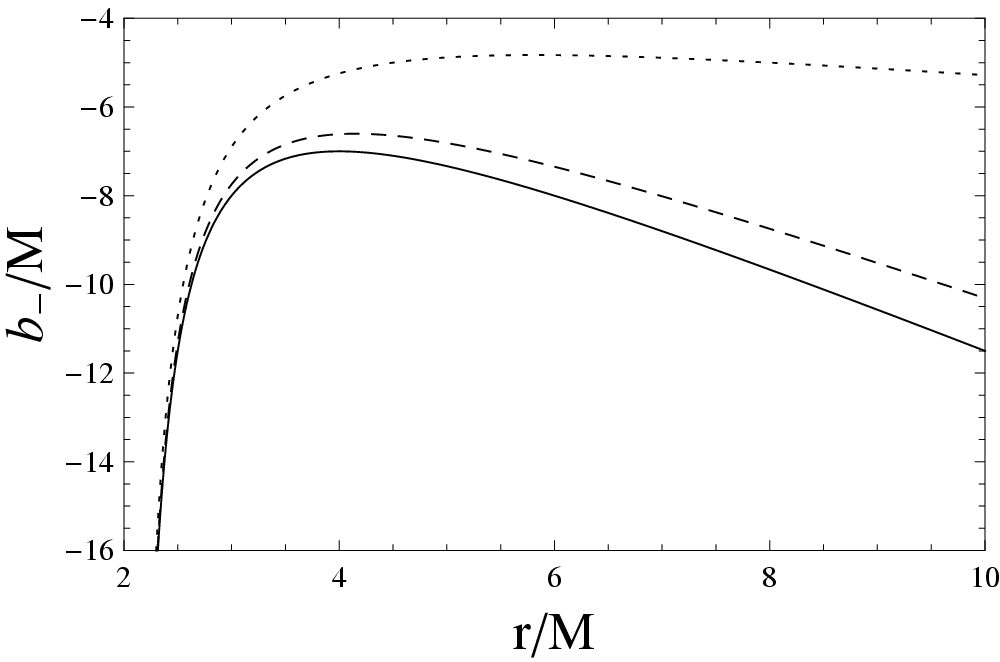}
  \caption{The radial dependence of impact parameters for the different values of $\kappa$.
  The solid, dashed, and dotted lines correspond to values of $j=0$, $j=0.5$, and $j=1$, respectively. }\label{turnpoints}
\end{figure*}

In the Fig.~\ref{turnpoints}
the radial dependence of
the impact parameter
$b_{\pm}$ for the different
values of the parameter
$\kappa$ are shown.
From the figure one can see that with increasing the value
of $\kappa$ the interval of impact parameter for
capturing the massless particles is decreased.

In the Table~\ref{table1}
the values of local maximum of $b_{-}$ are listed.
From the results one can see increase of the value of
$b_{-{\rm max}}$
with increase of parameter $\kappa$ which
corresponds to decreasing the interval of impact parameter for capturing massless particles.


Recall for massive particles
$\EE=E/m$,
$\LL=L/(mM)$, and
$\JJ=J/m$.
Solving equation $V(r)=0$ for massive particle one can obtain
\begin{eqnarray}
\LL &=&\LL_{\pm}(r)\label{momenta1}\\\nonumber&=&\frac{-2aM\EE \pm r \sqrt{\Delta(r)[\EE^{2}-(1-2M/r)(1+\JJ^2)]}}{M(r-2M)}.
\end{eqnarray}
Now one may conclude that a massive particle with angular momentum
$\LL =\LL_{\pm} (r)$
has a turning point at the position $r$.
For bound orbits,
i.e.
$\EE<1+\JJ^2$,
$V(r)$
is positive as radial coordinate increases to higher values,
showing
that they
are not able to
 reach infinity
 but will return back.
 Thus one may concentrate
 on soft bound and
unbound orbits
i.e.
$\EE\ge 1+\JJ^2$.

Consider the collision of particles
$1$
and
$2$
and
creation of particles
$3$
and
$4$.
The local conservation of four-momentum after
the collision is given by
\begin{eqnarray}
\label{conservation}
p_{1}^{\alpha}+
p_{2}^{\alpha}
=
p_{3}^{\alpha}+
p_{4}^{\alpha},
\end{eqnarray}
and if  $\alpha=t$ and $\alpha=\varphi$ one gets conservation
of energy and angular momentum during the reaction.

For the particles
$1$
and
$2$ the masses and energies are defined
iand
$m_{3}$,
$E_{3}$ and
$L_{3}$ are specified,
then one can determine
$m_{4}$,
$E_{4}$, and
$L_{4}$.
In fact $m_{4}$ can be expressed in terms of the
quantities of other three particles as follows
\begin{eqnarray}
m_{4}^{2}&=&-
p_{4\alpha}p_{4}^{\alpha}=-(p_{1}^{\alpha}+p_{2}^{\alpha}-p_{3}^{\alpha})
(p_{1\alpha}+p_{2\alpha}-p_{3\alpha}).
\end{eqnarray}

The CM energy of particles
$1$ and
$2$
near the event horizon for a
extreme rotating black hole has the following form
\begin{eqnarray}
\frac{E_{\rm cm}}{m} \approx
\sqrt{\frac{2(2\EE_{2}-{\LL}_{2})(2\EE_{1}-\sqrt{3\EE_{1}^{2}-1-\JJ_{1}^2})}{\epsilon}},
\end{eqnarray}
where we denote the radius of the collision point
as
$r_{c}=M/(1-\epsilon)$
and
$0<\epsilon\ll 1$.
For a critical particle, condition $\EE_{1}>{1}/\sqrt{3}$ must be satisfied.
As $\epsilon \to 0 $, the CM energy is diverging.

\begin{figure*}[t!]
\includegraphics[width=0.45\textwidth]{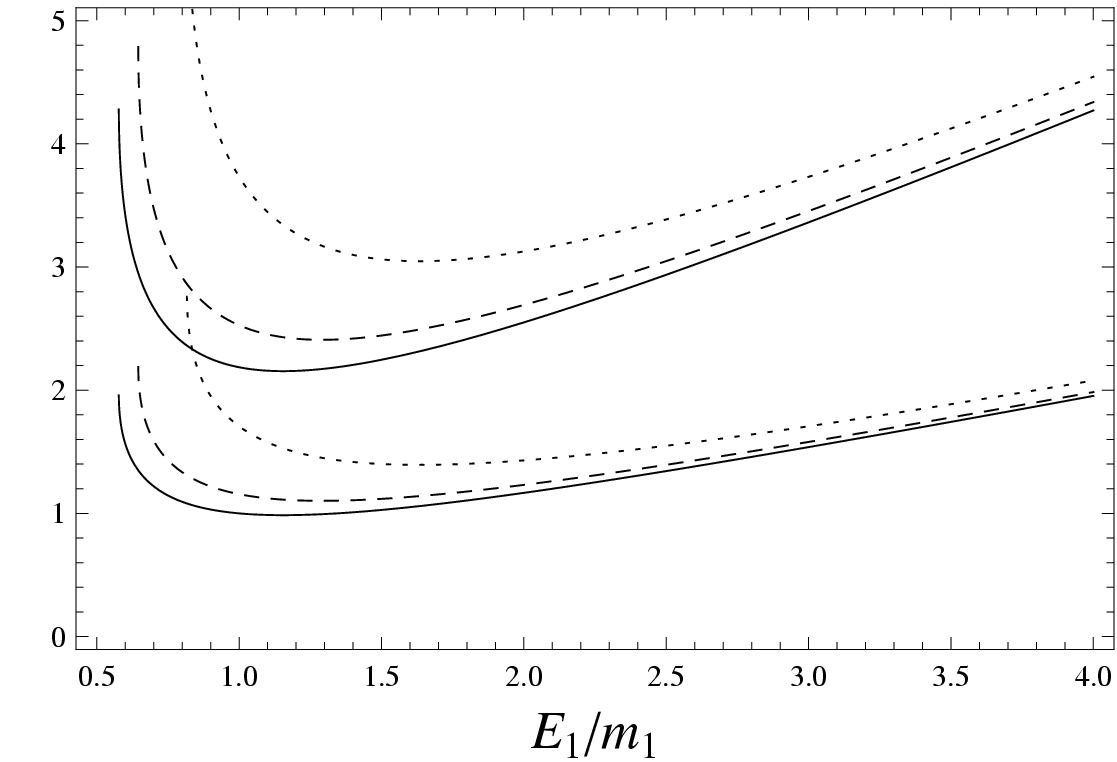}
\includegraphics[width=0.45\textwidth]{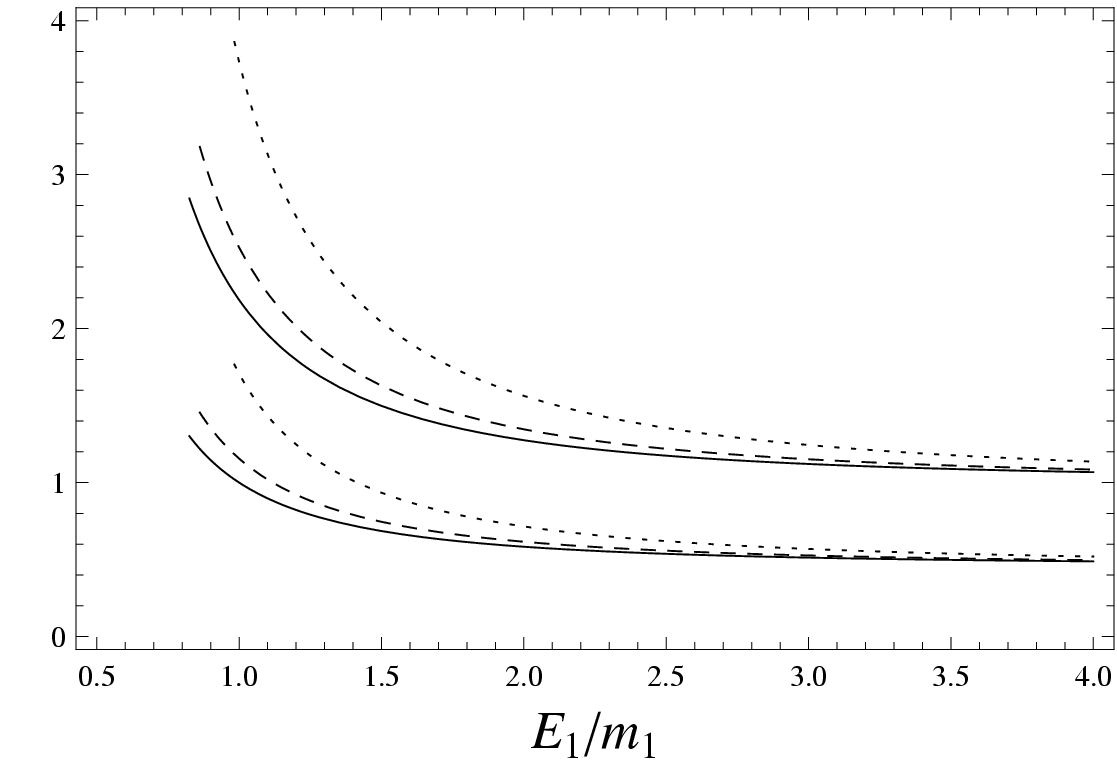}
\caption{\label{upends}
Upper limits of the mass and
energy
of the  emitted particle
as functions of the
energy of the
incident critical
particle.
The solid, dashed, and dotted lines correspond to the
values of $\JJ=0$, $\JJ=0.5$, and $\JJ=1$, respectively. }
\end{figure*}

Using the conservation
law~(\ref{conservation}) and considering
the particle
$3$
 as escaping, the upper limits of the mass and
energy of the emitted particle are given as
\begin{eqnarray}
m_{3}
\le (2E_{1}-\sqrt{3E_{1}^{2}-m_{1}^{2}-J_{1}^2})/(2-\sqrt{2})=
m_{B}\ \ \
\end{eqnarray}
and
\begin{equation}
E_{3}\le (2E_{1}-\sqrt{3E_{1}^{2}-m_{1}^{2}-J_{1}^2})/(2-\sqrt{3})=E_{B},
\end{equation}
respectively. Note that $\lambda_{+}=E_{B}$ can be realised only
if particle
$
3$
 is massless. Figure~\ref{upends} shows the
upper limits as function of $E_{1}/m_{1}$. One can easily see that
the presence of parameter $\JJ$ increases the upper limit of
energy of extracted particle.

Using the above result one can easily constrain the efficiency of
the  energy extraction. The unconditional upper limit is given by
\begin{eqnarray*}
\eta_{B}=
\frac{\lambda}{m_{1}+m_{2}} =
\frac{(\sqrt{2}+1)\lambda}{\sqrt{2}m_{1}+\lambda},
\end{eqnarray*}
where $\lambda=(2E_{1}-\sqrt{3E_{1}^{2}-m_{1}^{2}-J_{1}^2})/(2-\sqrt{3})$.

Since $\lambda$ takes the upper limit for $m_{3}=0$ the
unconditional  upper limit will depend on the value of $J_{1}$. In
the Table~\ref{table2} the dependence of energy extraction
efficiency for the different values of $\JJ$ is shown. One can see
that the maximal energy extraction can be  increased up to  $ 203.
\%$

\begin{table*}
\caption{\label{table2} The values of the energy extraction efficiency parameter. }
\begin{ruledtabular}
\begin{tabular}{|c|c|c|c|c|c|c|c|c|c|c|c|}
  $\JJ$ & 0 & 0.2 & 0.4 & 0.6 & 0.8 & 0.9 & 1 & 1.1 & 1.2 & 1.3 & 1.41 \\
  $\eta$& 1.46 & 1.48 & 1.52 & 1.58 & 1.66 & 1.70 & 1.75 & 1.80 & 1.85 & 1.91 & 2.03
\end{tabular}
\end{ruledtabular}

\end{table*}

\section{Conclusion\label{secVI}}

In this paper we have studied the effect of the high
center-of-mass energy of the collision of two test particles of
the same mass in the spacetimes of (i) the static black string,
(ii) rotating black string and (iii) the black string immersed in
external magnetic field.

The model considered here, the 4D Kerr spacetime with a flat
additional dimension, is not an exact solution of the Einstein
equations; therefore, a curved spacetime in the  additional
dimension has to be expected in realistic situations when the
spacetime is expected to be an exact solution of the Einstein
equations, or their modifications. However, this model can be
useful, at least, at the moment of collision of the particles and
can represent (locally) the outcome of the collision, since
locally the spacetime can be always approximated by the flat
geometry. Thus, one can describe the local influence of the
additional dimension, e.g., the leaking or appearance of
{unlimited} energy after the collision.

In particular we have shown that the BSW mechanism originally
investigated in $S^2$ topology can be {extended} to $S^2 \times
\mathbb{R}^1$ topology. We have shown that {due to} the particular
property of the motion in black string spacetime, namely, the
appearance of new integral of motion $J$, leads to increase of the
energy of the particle and as a result increases the maximal
center-of-mass energy of the collision of particles. We have
demonstrated that there exists a critical angular momentum of a
freely falling particle to achieve a horizon of black string and
found that when the extra dimension is added, the absolute values
of the critical angular momenta increase for any value of the spin
parameter $a$.

We have demonstrated that the center-of-mass energy can be high
for a collision of a particle falling from infinity with a charged
particle moving at ISCO. In fact, this energy formally infinitely
grows when the ISCO shifts arbitrarily close to the horizon which
is the result of the influence of the Lorentz force acting on the
charged particle. However this energy cannot be unlimited {for}
realistic magnetic fields since the {estimated} values of the
strength of the magnetic fields in relativistic astrophysics are
restricted by values $B\sim10^8 Gs$ for the stellar mass black
holes and $B\sim10^4 Gs$ for the supermassive black holes
\cite{Frolov:2012}.

It has been shown that the collisional energy of particles can be
ultrahigh not only for extremely rotating black string, but also
when the rotation of the black string is not maximal, $a<1$, if to
take into account the probability of the multiple scattering of a
freely falling particle with particles from the accretion disc. In
the calculation of such a process we used a simplified model where
the gravitational field of the accretion disc has been considered
infinitely small in comparison with the gravitation of the black
string and has not been taken into account.

The frame-dragging effects in a 4D Kerr black hole spacetime can
accelerate particles and one needs significant fine-tuning to get
sensible cross sections for particles (at least one of particles
has to have critical angular momentum). Here we show that in the
case  of black string we still have fine-tuning but the fine-tuned
value of critical angular momentum depends on a new constant of
motion appearing due to the extra dimension. The increase of
critical angular momentum in the background 5D black string
spacetime is accompanied by a decrease of stability of the
particle's trajectories.

We {have} also analyzed the energy extraction from rotating black
string. In the case of am extreme rotating black hole the energy
extraction efficiency has the upper limit $146\%$. We have shown
that the presence of the extra dimension can, in principle,
increase the upper limit of efficiency of energy extraction up to
$203\%$.

\ \

\begin{acknowledgments}

The authors would like to express their acknowledgements for the
institutional support of the Faculty of Philosophy and Science of
the Silesian University at Opava, the internal student grant of
the Silesian University SGS/23/2013 and EU grant Synergy
CZ.1.07/2.3.00/20.0071. A. A. and B. A. thank the TIFR, IUCAA
(India), and Max Planck Institut f$\rm{ \ddot{u}}$r
Gravitationsphysik (Albert Einstein Institute, Germany) for warm
hospitality. This research is supported in part by the projects
F2-FA-F113, FE2-FA-F134, F2-FA-F029 of the UzAS; {by the TWAS
associateship grants}; by the ICTP through the OEA-NET-76,
OEA-PRJ-29 projects; and the Volkswagen Stiftung (Grant No. 86
866).

\end{acknowledgments}


\begin{thebibliography}{References}


\bibitem{Bondarescu:2008}
  M.~Bondarescu
  ``Simple solutions to the Einstein Equations in
spaces with unusual topology,''
   hep-th/0504057 (2008)


\bibitem{Horowitz2002}
  G.~T.~Horowitz
  ``Playing with Black Strings,''
   hep-th/0205069 (2002).





\bibitem{Grunau:2013}
  S.~Grunau and B.~Khamesra
  ``Geodesic motion in the (rotating) black string spacetime,''
  arXiv:1303.6863 [gr-qc] (2013).



\bibitem{aliev1}
{A.N. Aliev, A.E. Gumrukcuoglu,
``Charged rotating black holes on a 3-brane,''
Physical Review D {\bf 71 }, 104027  (2005). }

\bibitem{schee1}
{J. Schee, Z. Stuchlik,
 ``Optical phenomena in the field of braneworld Kerr black holes,''
Int. Journal of Modern Physics D  {\bf 18 } 983- (2009)
}
\bibitem{zd80}{Z. Stuchlik,
  ``Equatorial circular orbits and the motion of the shell of dust in the field of a rotating naked singularity,''
  Bulletin of the Astronomical Institutes of Czecjoslovakia, {\bf
31}, 129 (1980).}



\bibitem{pk08}
S. Pal and S. Kar,
``Gravitational lensing in braneworld gravity: formalism and applications,''
Class. Quantum Grav., {\bf 25}, 045003    (2008)


\bibitem{aaprd}
{A. A. Abdujabbarov and B. J. Ahmedov,
``Test particle motion around a black hole in a braneworld,''
Phys. Rev. D \textbf{81},
044022 (2010)}

\bibitem{zdnk}
J. Schee and Z. Stuchlik, ``Profiles of emission lines generated
by rings orbiting braneworld Kerr black holes,'' Gen. Rel.
Gravit., {\bf 41}, 1795 (2009)

\bibitem{zdnkktrlv}
Z. Stuchlik and A. Kotrlova, ``Orbital resonances in discs around
braneworld Kerr black holes,'' Gen. Rel. Gravit., {\bf 41}, 1305
(2009)



\bibitem{lobo08}
{C.G. B\"{o}hmer, T. Harko, F.S.N. Lobo,
``Solar system tests of brane world models,''
Class. Quantum Grav.
{\bf 25}, 045015 (2008)}

\bibitem{pkh08}
{C.S.J. Pun, Z. Kov\'{a}cs, T. Harko,
``Thin accretion disks onto brane world black holes,''
 Phys. Rev. D {\bf 78},
084015 (2008)}

\bibitem{Lam08}
{E. Hackman, V. Kagramanova, J. Kunz, C. L\"{a}mmerzahl,
``Analytic solutions of the geodesic equation in higher dimensional static spherically symmetric spacetimes,''
Phys. Rev. D {\bf 78}, 124018 (2008)}



\bibitem{Aliev:1988wv}
  A.~N.~Aliev and D.~V.~Galtsov,
  ``Gravitational Effects in the Field of a Central Body Threaded by a Cosmic String,''
  Sov.\ Astron.\ Lett.\  {\bf 14}, 48 (1988).

\bibitem{Galtsov:1989ct}
  D.~V.~Galtsov and E.~Masar,
  ``Geodesics In Space-times Containing Cosmic Strings,''
  Class.\ Quant.\ Grav.\  {\bf 6}, 1313 (1989).

\bibitem{Chakraborty:1991mb}
  S.~Chakraborty and L.~Biswas,
  ``Motion of test particles in the gravitational field of cosmic strings in different situations,''
  Class.\ Quant.\ Grav.\  {\bf 13}, 2153 (1996).

\bibitem{Ozdemir:2003km}
  N.~Ozdemir,
  ``Gravitomagnetic effects and cosmic strings,''
  Class.\ Quant.\ Grav.\  {\bf 20}, 4409 (2003).

\bibitem{Ozdemir:2004ne}
  F.~Ozdemir, N.~Ozdemir and B.~T.~Kaynak,
  ``Multi-black holes solution with cosmic strings,''
  Int.\ J.\ Mod.\ Phys.\ A {\bf 19}, 1549 (2004).

\bibitem{Hackmann:2009rp}
  E.~Hackmann, B.~Hartmann, C.~L\"ammerzahl and P.~Sirimachan,
  ``The Complete set of solutions of the geodesic equations in the space-time of a Schwarzschild black hole pierced by a cosmic string,''
  Phys.\ Rev.\ D {\bf 81}, 064016 (2010)
  [arXiv:0912.2327 [gr-qc]].

\bibitem{Hackmann:2010ir}
  E.~Hackmann, B.~Hartmann, C.~L\"ammerzahl and P.~Sirimachan,
  ``Test particle motion in the space-time of a Kerr black hole pierced by a cosmic string,''
  Phys.\ Rev.\ D {\bf 82}, 044024 (2010)
  [arXiv:1006.1761 [gr-qc]].

\bibitem{Hartmann:2010rr}
  B.~Hartmann and P.~Sirimachan,
  ``Geodesic motion in the space-time of a cosmic string,''
  JHEP {\bf 1008}, 110 (2010)
  [arXiv:1007.0863 [gr-qc]].

\bibitem{Hartmann:2012pj}
  B.~Hartmann and V.~Kagramanova,
  ``Geodesic motion in the space-time of cosmic strings interacting via magnetic fields,''
  Phys.\ Rev.\ D {\bf 86}, 045028 (2012)
  [arXiv:1204.0396 [hep-th]].

\bibitem{Hartmann:2010vp}
  B.~Hartmann, C.~L\"ammerzahl and P.~Sirimachan,
  ``Detection of cosmic superstrings by geodesic test particle motion,''
  Phys.\ Rev.\ D {\bf 83}, 045027 (2011)
  [arXiv:1012.3285 [hep-th]].

\bibitem{BSW:2009}
 M.~Ba$\mathrm{\tilde{n}}$ados, J.~Silk and S.~M.~West,
 ``Kerr Black Holes as Particle Accelerators to Arbitrarily High Energy,''
 Phys.\ Rev.\ Lett.\ {\bf 103}, 111102 (2009)


\bibitem{Said-Adami:2011}
 J.~L.~Said and K.~Z.~Adami,
 ``Rotating charged cylindrical black holes as particle
 accelerators,''
 Phys.\ Rev.\ D\ {\bf 83},  104047 (2011)

\bibitem{Jac-Sot:2010}
 T.~Jacobson and T.~P.~Sotiriou,
 ``Spinning Black Holes as Particle Accelerators,''
 Phys.\ Rev.\ Lett.\ {\bf 104},  021101 (2010)

\bibitem{Zaslavsky:2012}
 O.~B.~Zaslavsky,
 ``Acceleration of particles by rotating black holes: near-horizon geometry and kinematics,''
 Grav. Cosmol. {\bf 18}, 139 (2012)


\bibitem{Frolov:2010}
 V.~P.~Frolov and A.~A.~Shoom,
 ``Motion of charged particles near weakly magnetized Schwarzschild black hole,''
 Phys. Rev. D {\bf 82}, 084034 (2010)

\bibitem{wald}R.M. Wald,
``Black hole in a uniform magnetic field,''
Phys. Rev. D \textbf{10}, 1680 (1974).

\bibitem{Frolov:2012}
 V.~P.~Frolov,
 ``Weakly magnetized black holes as particle accelerators,''
 Phys. Rev. D {\bf 85}, 024020 (2012)

\bibitem{Abd-Tur-Ahm-Kuv:2013}
 A.~A.~Abdujabbarov, A.~A.~Tursunov, B.~J.~Ahmedov and A.~Kuvatov,
 ``Acceleration of particles by black hole with gravitomagnetic
charge immersed in magnetic field,''
 Astrophys.\ Space.\ Sci.\ {\bf 343}, 173 (2013)

\bibitem{Abd-Ahm-Jur:2013}
 A.~A.~Abdujabbarov, B.~J.~Ahmedov and N.~B.~Jurayeva,
 ``Charged-particle motion around a rotating non-Kerr black hole immersed in a uniform magnetic field,''
 Phys. Rev. D  {\bf 87}, 064042 (2013)

\bibitem{Pat-Josh:CQG:2011}
 M.~Patil and P.~Joshi, ``Kerr naked singularities as particle
 accelerators,'' Class. Quantum Grav. {\bf 28} 235012 (2011)










\bibitem{Stu-Hle-Tru:CQG:2011}
 Z.~Stuchl{\'i}k, S.~Hled{\'i}k and K.~Truparov{\'a},
 ``Evolution of Kerr superspinars due to accretion counterrotating
Keplerian discs,'' Class. Quantum Grav. {\bf 28} 155017 (2011)


\bibitem{Stuchlik-Schee:CQG:2012-1}
 Z.~Stuchl{\'i}k and J.~Schee,
 ``Observational phenomena related to primordial Kerr superspinars,''
 Class. Quantum Grav. {\bf 29}  065002 (2012)


\bibitem{Stuchlik-Schee:CQG:2012-2}
 Z.~Stuchl{\'i}k and J.~Schee,
 ``Counter-rotating Keplerian discs around Kerr superspinars,''
 Class. Quantum Grav. {\bf 29}  025008 (2012)

\bibitem{Stuchlik-Schee:CQG:2013}
 Z.~Stuchl{\'i}k and J.~Schee, ``Ultra-high-energy collisions in the superspinning Kerr
geometry,'' Class. Quantum Grav. {\bf 30} 075012 (2013)

\bibitem{stuchlik_10}{Z.~Stuchl{\'i}k and J.~Schee,
``Appearance of Keplerian discs orbiting Kerr superspinars,''
Class. Quantum Grav. {\bf 27}, 215017 (2010).}




\bibitem{Ata-Ahm-Sha:2013}
 F.~Atamurotov, B.~Ahmedov and S.~Shaymatov,
 ``Formation of black holes through BSW effect and black
hole–black hole collisions,''
 Astrophys.\ Space.\ Sci.\ {\bf 347}, 277 (2013)



\bibitem{takami}{
L. Rezzolla, K. Takami,
``Black-hole production from ultrarelativistic collisions,''
Class. Quantum Grav., {\bf 30}, 012001 (2013). }

\bibitem{japan1}{
T. Harada, M. Kimura,
``Collision of an innermost stable circular orbit particle around a Kerr black hole,''
Physical Review D, {\bf 83},  024002 (2011).}

\bibitem{japan2}{T. Harada, M. Kimura,
``Collision of two general geodesic particles around a Kerr black hole'
Physical Review D, {\bf 83},   084041 (2011).}

\bibitem{japan3}{N. Tsukamoto, M. Kimura, T. Harada,
``High Energy Collision of Particles in the Vicinity of Extremal Black Holes in Higher Dimensions: Banados-Silk-West Process as Linear Instability of Extremal Black Holes,'
arXiv:1310.5716 (2013).}

\bibitem{japan4}{
H. Nemoto, U. Miyamoto, T. Harada, T. Kokubu,
``Escape of superheavy and highly energetic particles produced by particle collisions near maximally charged black holes,'
Physical Review D, {\bf 87}, 127502 (2013).}

\bibitem{japan5}{T. Harada, H. Nemoto, U. Miyamoto,
``Upper limits of particle emission from high-energy collision and reaction near a maximally rotating Kerr black hole,'
Physical Review D, {\bf  86},  024027 (2012).
}

\bibitem{dadhich}{A. Abdujabbarov,  N. Dadhich, B. Ahmedov, H. Eshkuvatov,
``Particle acceleration around a five-dimensional Kerr black hole,'
Physical Review D {\bf 88}, 084036 (2013).}


\bibitem{ONeill:1995}
 B.~O'Neill, ``The Geometry of Kerr Black Holes'', AK Peters, Wellesley, Massachusetts, (1995).






\bibitem{Sij-Chan:2011:PRD}
  S.~Gao and C.~Zhong,
  ``Non-extremal Kerr black holes as particle accelerators,''
  Phys.\ Rev.\ D {\bf 84}, 044006 (2011)
  [arXiv:1106.2852 [gr-qc]].



\bibitem{Grib-Pavlov:2011:Grav-Cos}
  A.~A.~Grib and Yu.~V.~Pavlov,
  ``On particle collisions near rotating black holes,''
  Grav. and Cosmol. {\bf 17}, 42 (2011)



\bibitem{Thorne:1974}
  K.~S.~Thorne, "Disk Accretion onto a Black Hole. II. Evolution of the Hole," Astrophys. J. {\bf 191}, 507 (1974).
\end{thebibliography}
\end{document}